\documentclass[prd,twocolumn,nofootinbib,showpacs]{revtex4}

\usepackage{amsmath}
\usepackage{graphicx}
\usepackage{hyperref}

\newcommand{\cpp}[1]{\texttt{#1}}

\newcommand{\vect}[1]{\vec{#1}}
\newcommand{\tens}[1]{\mathsf{#1}}

\newcommand{\onehalf}{{\textstyle \frac{1}{2}}}

\newcommand{\mylistitem}[1]{\vspace{6pt}\noindent\textbf{#1}}
\newcommand{\lastlistitem}{\vspace{6pt}}

\begin{document}

\title{Synthetic LISA: Simulating Time Delay Interferometry in a Model LISA}


\author{Michele Vallisneri}

\affiliation{Jet Propulsion Laboratory, California Institute of Technology, Pasadena, CA 91109}
\email{Michele.Vallisneri@jpl.nasa.gov}

\begin{abstract}
We report on three numerical experiments on the implementation of Time-Delay Interferometry (TDI) for LISA, performed with \emph{Synthetic LISA}, a C++/Python package that we developed to simulate the LISA science process at the level of scientific and technical requirements. Specifically, we study the laser-noise residuals left by first-generation TDI when the LISA armlengths have a realistic time dependence; we characterize the armlength-measurements accuracies that are needed to have effective laser-noise cancellation in both first- and second-generation TDI; and 
we estimate the quantization and telemetry bitdepth needed for the phase measurements.
\emph{Synthetic LISA} generates synthetic time series of the LISA fundamental noises, as filtered through all the TDI observables; it also provides a streamlined module to compute the TDI responses to gravitational waves according to a full model of TDI, including the motion of the LISA array and the temporal and directional dependence of the armlengths. We discuss the theoretical model that underlies the simulation, its implementation, and its use in future investigations on system characterization and data-analysis prototyping for LISA.
\end{abstract}


\maketitle

\section{Introduction}
\label{sec:intro}

The Laser Interferometer Space Antenna (LISA) is a joint NASA--ESA deep-space mission aimed at detecting and studying gravitational radiation in the $10^{-5}$--$10^{-1}$ Hz frequency band \cite{PPA98}. It is expected to be launched in the year 2012, and to start collecting scientific data approximately a year later, after reaching its final orbital configuration \cite{Folkner97}. LISA consists of three widely separated spacecraft flying in a triangular, almost equilateral configuration, and exchanging coherent laser beams; gravitational waves (GWs) will be measured by picometer interferometry as modulations in the distance between the spacecraft.

LISA, which will operate in a lower frequency band than ground-based GW interferometers, holds the promise of providing access to entirely new classes of GW sources, but it also introduces complications unknown to ground-based detectors, such as the complex signal and noise transfer functions, the problem of canceling the otherwise overwhelming laser phase noise in an unequal-arm interferometer, the necessity of dealing simultaneously with many continuous signals (including a confusion-noise background of galactic white-dwarf binaries), and the possibility of using multiple interferometric observables as a virtual network of GW interferometers. These complications hinder the analytical characterization of LISA's detection capabilities as a function of its configuration, as well as the development of data-analysis techniques aimed at specific GW sources. Computer simulations will therefore play a crucial role in exploring LISA's performance, in obtaining insight about its optimal operation, and in prototyping and testing data-analysis protocols.

In this paper we report on three numerical investigations performed with \emph{Synthetic LISA}, a software package that we developed at the Jet Propulsion Laboratory to simulate the LISA science process at the interface of scientific and technical mission requirements. All investigations focus on outstanding implementation issues for Time-Delay Interferometry (TDI), the LISA-specific technique currently envisaged to suppress the otherwise overwhelming laser phase noise by combining (with delays) the basic LISA phase measurements aboard the three spacecraft into composite laser-noise--free observables (see Sec.\ \ref{sec:tdi} for a discussion and full  references). More in detail:
\begin{enumerate}
\item In Sec.\ \ref{sec:expsecondgen}, we give the first quantitative estimate, based on a straight simulation, of the improvement in laser-noise stabilization that would eliminate the need for second-generation TDI for a realistic flexing LISA array \cite{flexy2003,STEA2003,TEA2004} using standard Michelson observables. We find that an rms improvement factor between 3 and 10 is sufficient. We give also numerical evidence of effective laser-noise subtraction with second-generation observables.
\item In Sec.\ \ref{sec:exparmlengths}, we evaluate the armlength-ranging accuracy \cite{TSSA2003} that would be required for effective laser-noise cancellation in first- and second-generation TDI Michelson observables. We find that ranging accuracies between 30 m and 100 m (rms) are adequate when simple linear extrapolation is used to compute the armlengths between measurements.
\item In Sec.\ \ref{sec:quantize}, we estimate the granularity that can be allowed in the quantization of phase measurements while preserving effective laser-noise cancellation. Assuming white laser frequency noise bandlimited at 1 Hz, we find that a total of 32--34 (or 36--38) bits per sample are needed for the Michelson observables of first-generation (second-generation) TDI. 
\end{enumerate}
We present these results as representative of the numerical experiments that become possible with state-of-the-art simulators such as \emph{Synthetic LISA}, and we suggest possible directions of investigation in the final section of this paper.

\emph{Synthetic LISA} represents the evolution of previous simulation tools developed in the LISA Project \cite{lisasimwhitepaper}. Among other improvements, \emph{Synthetic LISA} is based on a complete model of TDI: the LISA armlengths  change realistically with the motion of the array; the laser beams propagate causally; and a full set of TDI combinations can be generated. \emph{Synthetic LISA} joins other existing software that simulates the LISA response to noise and GWs, such as the well-established \emph{LISA Simulator} by N.\ Cornish and L.\ Rubbo \cite{cornishrubbo}. Why write a new simulator, then? Being able to rely on a plurality of simulation tools allows for mutual validation and verification, which is crucial if implementation choices must be predicated on the results of numerical experiments. In addition, the two simulators have a slightly different focus. The \emph{LISA Simulator} was conceived to interface source simulations to data analysis, while \emph{Synthetic LISA} was targeted to explore the interaction between LISA science and technology, and it must therefore operate at a lower level of abstraction: in particular, \emph{Synthetic LISA} performs an explicit time-domain simulation of interferometry, including the cancellation of laser phase noise. On the other hand, it operates at a higher level of abstraction than integrated-modeling simulations \cite{merkowitz}: it does not need to model spacecraft subsystems, but rather it assumes nominal specifications of their performance.

This paper is laid out as follows. In Sec.\ \ref{sec:modeling} we describe the theoretical model of the LISA science process used in our simulations; in Sec.\ \ref{sec:synthlisa} we briefly discuss the implementation and usage of \emph{Synthetic LISA}; in Sec.\ \ref{sec:tests} we report on our main numerical experiments; and in Sec.\ \ref{sec:conclusions} we give our conclusions.
Appendices \ref{app:conventions} and \ref{app:synthnoise} describe, respectively, the geometric conventions and the treatment of noise used in \emph{Synthetic LISA}.  
In the following, we set $G = c = 1$ unless otherwise indicated.

\section{Modeling of a Synthetic LISA}
\label{sec:modeling}
\begin{figure}\begin{center}
\includegraphics[width=3.2in]{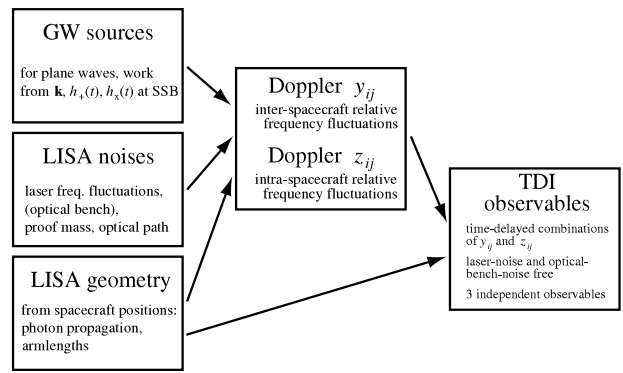}
\end{center}\caption{A block diagram of the LISA science process.
\label{fig:lisablock}}
\end{figure}

Figure \ref{fig:lisablock} is a block diagram of the LISA science process, as modeled in \emph{Synthetic LISA}. At the top of the hierarchy sit the TDI observables, which represent the main scientific product of the mission, and which will be run through data-analysis algorithms to search for GW signals. The TDI observables are time-delayed combinations of the basic interferometric measurements ($y$ and $z$) that compare the frequencies (or phases) of the two lasers on each spacecraft between themselves, and with the lasers incoming from the other two spacecraft. The Doppler measurements bear the imprint of the instrumental noises and of the GW signals, but the latter can be read off efficiently only from the TDI observables, which are free of the otherwise overwhelming laser phase noise and optical-bench noise. The time-dependent geometry of laser propagation across the LISA array influences the effect of the LISA noises and (especially) of GW signals on the Doppler measurements; a precise knowledge of geometry is needed also to build the TDI observables in such a way that laser phase noise and optical-bench noise are canceled effectively. In this section we go through all the elements of Fig.\ \ref{fig:lisablock}, and discuss in detail how they are modeled in \emph{Synthetic LISA}. In Sec.\ \ref{sec:lisamodel} we describe the geometry of the LISA array, and the setup of the interferometric payload on each spacecraft; in Secs.\ \ref{sec:gwresponse} and \ref{sec:noiseresponse} we describe the response of the basic interferometric observables to GWs and to the LISA fundamental noises; last, in Sec.\ \ref{sec:tdi} we give a rapid overview of TDI as used in LISA.

\subsection{LISA geometry and interferometry}
\label{sec:lisamodel}

The motion of the LISA array is complex: at the qualitative level, the three LISA spacecraft maintain a quasi-equilateral triangular configuration (where the arms stay equal to about 1\%) trailing the Earth along its orbit in the plane of the ecliptic; at the same time, the constellation maintains an inclination of $\pi/2 - \pi/6 = \pi/3$ with respect to the plane of the ecliptic (as measured from the normal of the instantaneous plane of the LISA constellation to the normal to the plane of the ecliptic), and it performs a cartwheeling motion, rotating around the normal to the instantaneous LISA plane with a rotation period of a year. This picture is realized in practice by placing the three spacecraft on eccentric, inclined solar orbits \cite{Folkner97}.

This pattern of motion improves the sensitivity of LISA to GW signals, making it more homogeneous over the sky (because the dependence of the antenna patterns to source position is averaged during the year), and improving the estimation of source position and polarization (because the GW responses become modulated by the variation of the antenna patterns). This added sensitivity comes at the price of complicating the GW response: the modulations induced by the changing orientation of the LISA plane spread the power of originally monochromatic GW signals, generating several sidebands at frequency multiples of $1/\mathrm{yr}$ \cite{giampieri}; furthermore, the relative motion of the detector with respect to the GW source introduces a time-dependent Doppler shift, which is the dominant effect for signals above $10^{-3}$ Hz [the characteristic bandwidth of the Doppler shifting is $\sim (\Omega R/c)f$, where $f$ is the GW frequency and $\Omega = 2 \pi / \mathrm{yr}$ is the LISA orbital angular velocity].
\begin{figure}\begin{center}
\includegraphics[width=3.2in]{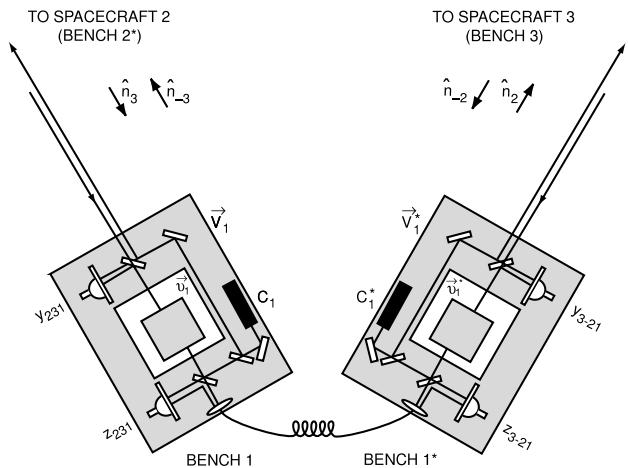}
\end{center}\caption{Schematic diagram of the proof-mass and optical-bench assemblies within each LISA spacecraft (adapted from Ref.\ \cite{PRDWB00}).}
\label{fig:assembly}
\end{figure}

When LISA is in operation, each spacecraft will exchange laser beams with the other two, measuring the phase of the arriving laser beams with respect to the local lasers; the laser beams are bounced off freely-falling proof masses that are shielded by the spacecraft from most external disturbances,\footnote{For instance, the spacecraft cannot completely shield the proof masses from cosmic rays.} so that they can serve as references for the measurement of GWs. To implement this measurement scheme, each spacecraft will carry two lasers, two proof masses, and two optical-readout schemes. Figure \ref{fig:assembly} presents a schematic diagram (adapted from Ref.\ \cite{PRDWB00}) of the proof-mass and optical-bench assemblies within one of the LISA spacecraft, labeled ``1''; the other two spacecraft have identical setups. In short:
\begin{enumerate}
\item the left-hand bench receives the laser beam from spacecraft 2, bounces it off its proof mass, and compares it with the local laser (without bouncing the latter) at the upper photodetector;
\item via an optical fiber, the left-hand bench receives the right-hand--bench laser and compares it with the local laser (without bouncing the latter) at the lower photodetector;
\item the left bench sends out the local laser (without bouncing it) to spacecraft 2, and (after bouncing it off the other side of its proof mass) to the right-hand bench.
\end{enumerate}
The operation of the right-hand bench (and indeed, of the benches on the other two spacecraft) is similar. [A recent candidate redesign of the optical benches \cite{bonny} would implement the comparison of the two lasers on the two benches of the same spacecraft by measuring their phases separately, doing away with the optical fiber, and then subtracting the measurements. For the purpose of obtaining the laser-noise--free TDI signals (see Sec.\ \ref{sec:tdi}) this modification amounts only to a redefinition of the intra-spacecraft phase measurements \cite{aetprivate}, so in this paper, and indeed in \emph{Synthetic LISA}, we refer to the older architecture.]

In this setup, the physical observable of interest is the comparison of phase between the local laser and the incoming laser, which carries information about the variations induced by GWs in the inter-spacecraft optical path. The phase fluctuations of the lasers, however, are much larger than the GW-induced phase shifts, and must be subtracted before GWs can be resolved. In the last few years, a number of authors collaborated to develop a scheme (\emph{Time-Delay Interferometry}, or TDI) to subtract laser noise by carefully combining time-shifted series of the inter- and intra-spacecraft phase measurements; if the lasers are not phase-locked to a master (see the end of Sec.\ \ref{sec:tdi}), the intra-spacecraft phase measurements carry no information about GWs, but they do carry a combination of the phase noises from the lasers within each spacecraft.
\begin{figure}\begin{center}
\includegraphics[width=3.0in]{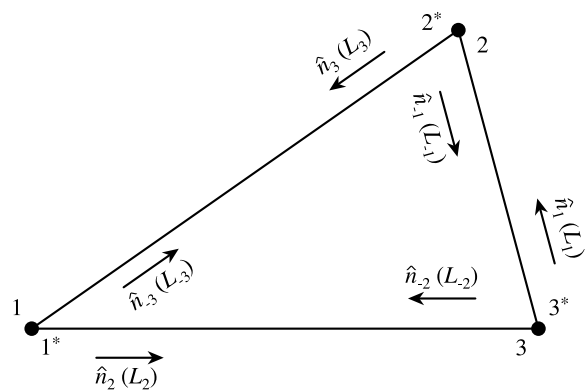}
\end{center}\caption{Schematic LISA configuration.  The spacecraft are labeled 1, 2, and 3; each spacecraft contains two optical benches, denoted by 1, $1^*$, 2, $2^*$, $3^*$, as indicated. The unit vectors $\hat{n}_l$ and light-path lengths $L_l$ connecting spacecraft $s$ and $r$ are indexed by $l=\{1,2,3\}$ for $(s,r) = \{(3,2),(1,3),(2,1)\}$, and $l=\{-1,-2,-3\}$ for $(s,r) = \{(2,3),(3,1),(1,2)\}$.\label{fig:geometry}}
\end{figure}

Because TDI has its origin in the techniques used to measure GWs by the Doppler tracking of distant spacecraft \cite{estawahl,tracking}, it prefers to describe the comparisons between laser beams in terms of fractional frequency differences rather than relative phase shifts (the two descriptions are exactly equivalent \cite{TEA2002}, as they are related by time integration). Thus, TDI represents the LISA readouts as \emph{basic Doppler observables}: $y_{slr}(t)$ is the fractional frequency difference at time $t$ between the beam received at spacecraft $r$(eceiver) from spacecraft $s$(ender) and the local laser; and $z_{slr}(t)$ is the analogous intra-spacecraft measurement \emph{on the same optical bench} (thus, although it carries the index $s$ it is in fact the fractional frequency difference between the two lasers of spacecraft $r$). In this paper, the index $l$(ink) denotes the (oriented) LISA arm along which the laser was transmitted, according to the cyclical indexing $l=\{1,2,3\}$ for $(s,r) = \{(3,2),(1,3),(2,1)\}$, and $l=\{-1,-2,-3\}$ for $(s,r) = \{(2,3),(3,1),(1,2)\}$ (thus, $\mathrm{sgn}(l) = \epsilon_{s|l|r}$). This spacecraft and link indexing is shown also in Fig.\ \ref{fig:geometry}. Note that our notation for the basic Doppler observables merges the two notations used in the scientific literature on \emph{first-generation} TDI ($y_{lr}$ and $z_{lr}$) and, more recently, on \emph{second-generation} TDI ($y_{sr}$ and $z_{sr}$).  Table \ref{tab:rosetta} shows a comparison (as it were, a Rosetta stone) of the notations used in various papers on TDI.
In the next two sections we discuss the response of these basic Doppler observables to GWs and to the noise sources present within each spacecraft.
\begin{table*}
\begin{center}
\includegraphics[width=\textwidth]{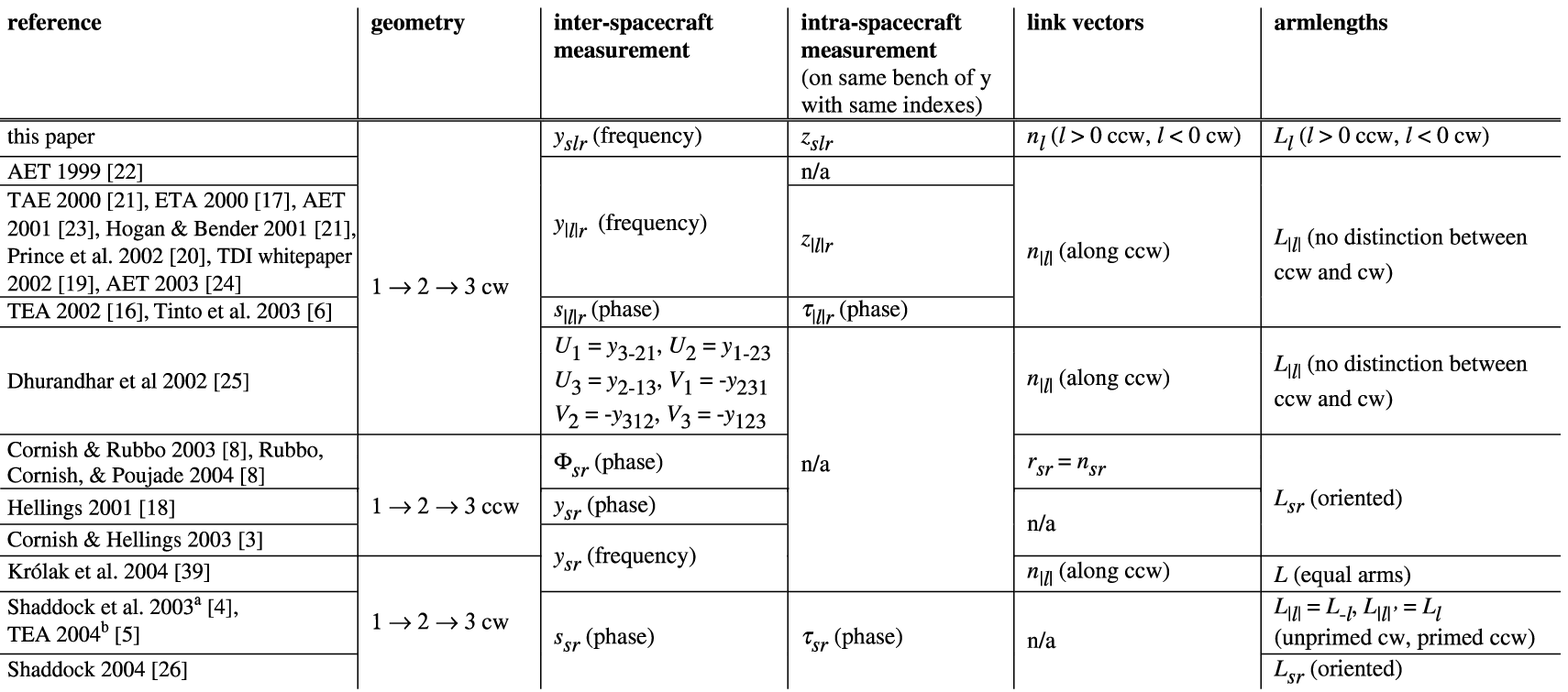}
\end{center}
\caption{A comparison of the phase-measurement and LISA geometry conventions used in the literature on TDI.
In the cited references, A, E, and T refer to J.\ W.\ Armstrong, F.\ B.\ Estabrook, and M.\ Tinto. Notations are described with respect to the usage of this paper, with $s \equiv$ sending spacecraft, $l \equiv$ armlink, $r \equiv$ receiving spacecraft; ``cw'' and ``ccw'' refer to the progression of spacecraft or link indexes, as seen when looking at the LISA constellation from above (from ecliptic latitude $90^\circ$ N); when indexes are shown in absolute values, only positive values are used. Tinto and Armstrong 1999 (not included in this table) has $y_1 \equiv$ two-way ccw ($a \rightarrow b \rightarrow a$), $y_2 \equiv$ two-way cw ($a \rightarrow c \rightarrow a$).
See \protect\url{www.vallis.org/tdi} for updates to this table.
\\ ${}^a$ The semicolon ordered-delay notation was introduced in Shaddock et al.\ 2003 and TEA 2004.
\\ ${}^b$ TEA 2004 uses $n_{|l|}$ to denote link vectors; it is ambiguous from the context whether these are ccw or cw.
\label{tab:rosetta}}
\end{table*}

\subsection{LISA response to gravitational waves}
\label{sec:gwresponse}

In this section we give an expression for the GW response of the basic Doppler observables $y_{ij}(t)$. Working in an inertial reference frame filled by a plane GW with propagation vector $\vect{k}$ and transverse--traceless gravitational tensor $\tens{h}(\vect{x};t) = \tens{h}(0,t-\vect{k}\cdot\vect{x}) \equiv \tens{h}(t)$, we denote the positions of the three spacecraft by $\vect{p}_i(t)$.
Following Estabrook and Wahlquist \cite{estawahl}, we write the response of the inter-spacecraft LISA Doppler observable $y_{slr}(t)$ to the plane wave as
\begin{multline}
y^\mathrm{gw}_{slr}(t) = \bigl[ 1 + \hat{k} \cdot \hat{n}_l(t) \bigr]
\times \bigl(
\Psi_l[t_\mathrm{send} - \hat{k} \cdot \vect{p}_s(t_\mathrm{send}) ] \\
- \Psi_l[t - \hat{k} \cdot \vect{p}_r(t) ]
\bigr),
\label{eq:responses}
\end{multline}
where $t_\mathrm{send}$ and $\vect{p}_s(t_\mathrm{send})$ are determined by the light-propagation equation $t_\mathrm{send} = t - |\vect{p}_r(t) - \vect{p}_s(t_\mathrm{send})|$, where $\hat{n}_l(t)$ is the oriented photon-propagation unit vector $\hat{n}_l(t) \propto \vect{p}_r(t) - \vect{p}_s(t_\mathrm{send})$, and where
\begin{equation}
\label{eq:bigpsi}
\Psi_l(t') = \frac{
\hat{n}_l(t) \cdot \tens{h}(t') \cdot
\hat{n}_l(t)}{2\bigl(1 - [\hat{k}\cdot\hat{n}_l(t)]^2\bigr)}.
\end{equation}
Equation \eqref{eq:responses} gives the inter-spacecraft Doppler observable for laser-beam reception at time $t$ on spacecraft $r$ from spacecraft $s$, through link $l$. The two $\hat{k} \cdot \vect{p}$ products correspond to the retardation of the plane wavefront to the position of the two spacecraft, while the $\hat{k} \cdot \hat{n}$ products come in as geometrical projection factors \cite{estawahl}. Equation \eqref{eq:responses} is not singular for $\hat{k} = \pm \hat{n}_l$, because in that case the transverse--traceless tensor $\tens{h}$ is orthogonal to $\hat{k}$ and $\hat{n}_l$, and the $\Psi_l$ go to zero.

The light-propagation equation defines the effective armlength $L_l(t)$ experienced by light propagating from $s$ to $r$, for reception at time $t$:
\begin{equation}
\label{eq:lightprop}
L_l(t) = |\vec{p}_\mathrm{r}(t) - \vec{p}_\mathrm{s}\bigl(t - L_l(t)\bigr)|.
\end{equation}
Note that in general $L_l(t) \neq L_{-l}(t)$.

The response to GWs of the intra-spacecraft Doppler observable $z_{slr}(t)$ is null, because the distance traveled by the intra-spacecraft beam is negligible for the GW amplitudes and wavelengths relevant to LISA.\footnote{If the lasers are not phase-locked to a master (see the end of Sec.\ \ref{sec:tdi}).\label{fn:phaselocking}}

\subsection{LISA response to fundamental noises}
\label{sec:noiseresponse}

In this section we give the response of the basic Doppler observables to the fundamental noise sources present within each spacecraft.  Looking back to Fig.\ \ref{fig:assembly}, we label the left-hand and right-hand optical benches (and their lasers) as $1$ and $1^*$, respectively (more generally, unstarred benches transmit into oriented arms with negative indices). Following Estabrook and colleagues \cite{ETA2000}, we denote the fractional frequency fluctuations of the laser on optical bench 1 as $C_1(t)$; these enter additively in the $y_{231}$ measurement,
together with the frequency noise from the laser on bench $2^*$ of spacecraft 2, retarded to the time of emission:
\begin{equation}
y^\mathrm{noise}_{231}(t) = C^*_{2}\bigl(t - L_3(t)\bigr) - C_1(t) + \cdots;
\end{equation}
next, the $y_{231}$ measurement is subject to noise due to fluctuations on the optical path of the beam incoming from spacecraft 2 (a combination of shot noise, pointing noise, and other optical-path noises), which we denote as $y^\mathrm{op}_{231}$; also, the velocity noise $\vec{v}_1$ of the proof mass on optical bench 1 (i.e., its deviation from perfect free fall) induces an additional Doppler shift on the incoming beam (the local beam does not bring in any velocity noise, since it is not bounced on the local proof mass):
\begin{equation}
y^\mathrm{noise}_{231}(t) = C^*_{2}\bigl(t - L_3(t)\bigr) - C_1(t) + y^\mathrm{op}_{231}(t) - 2 \vec{v}_1(t) \cdot \hat{n}_3(t) + \cdots;
\end{equation}
last, the random velocities $V^*_2$ and $V_1$ of the emitting and receiving optical benches (which are several orders of magnitude greater than $\vec{v}_1$) induce additional Doppler shifts with the same temporal structure of laser frequency noise:
\begin{multline}
y^\mathrm{noise}_{231}(t) = C^*_{2}\bigl(t - L_3(t)\bigr) - C_1(t) + y^\mathrm{op}_{231}(t) - 2 \vec{v}_1(t) \cdot \hat{n}_3(t) \\
+ \vec{V}^*_{2}\bigl(t - L_3(t)\bigr) \cdot \hat{n}_3(t) -
\vec{V}_{1}(t) \cdot \hat{n}_{-3}(t).
\label{eq:lasty}
\end{multline}
Along similar lines we derive the noise response of the intra-spacecraft measurement $z_{3-21}$ on spacecraft 1, which contains the frequency noises from lasers $1$ and $1^*$ at time $t$, the random velocities of the optical bench $1$ and of its proof mass, and the frequency shift $\eta_1$ upon transmission through the optical fiber (ultimately due to a component of the relative bench motions, $\vect{V}_1 - \vect{V}^*_1$):
\begin{equation}
z^\mathrm{noise}_{3-21}(t) = C_1(t) - C^*_1(t) + 2 \hat{n}_3(t) \cdot \vec{v}_1 + 
2 \hat{n}_{-3}(t) \cdot \vec{V}_1 + \eta_1;
\label{eq:lastz}
\end{equation}
here we are ignoring time-delay effects along the fibers.

Throughout the rest of this paper (and indeed, always in \emph{Synthetic LISA}) we take the optical-fiber noises and the optical-bench motions to be negligible. In fact, optical-fiber noise is removed in TDI by always using the $z_{slr}$ observables in pairs such as $(z_{231} - z_{3-21})/2$, $(z_{312} - z_{1-32})/2$, and so on.
One sees also that the optical-bench motions along the lines of sight (e.g., $\hat{n}_{-3} \cdot \vec{V}_1$, $\hat{n}_{2} \cdot \vec{V}^*_1$, and $\hat{n}_{3} \cdot \vec{V}^*_2$) can be absorbed in the corresponding laser frequency noise variables (e.g., $C_1$, $C_1^*$, and $C_2^*$), because they are appear in $y^\mathrm{noise}_{slr}$ and $z^\mathrm{noise}_{slr}$ with the same indices and the same evaluation times. Thus, if the TDI observables can successfully subtract laser frequency noise, they will also subtract the optical-bench motions, which are generally several orders of magnitude smaller.

In writing Eqs.\ \eqref{eq:responses}, \eqref{eq:lasty}, and \eqref{eq:lastz}, we have neglected also the offsets (up to several hundred MHz) between the center frequencies of the six LISA lasers, as well as the slow Doppler drifts resulting from the relative motion of the spacecraft (up to tens of MHz).  In practice, the frequency offsets and Doppler drifts will be corrected by down-converting the photodetector output and tracking fringe rates using onboard ultrastable oscillators (USOs) \cite{hell2001,TEA2002,TSSA2003}.  Although USOs introduce an important additional source of phase noise, their treatment is cumbersome, and we leave their modeling to a future version of \emph{Synthetic LISA}.

Under these assumptions, the simulation of the LISA noise response requires time series for 18 fundamental noise variables: the six proof-mass velocity noises along the line of sight (which we denote as $pm_1 \equiv \hat{n}_3 \cdot \vec{v}_1$, $pm_2 \equiv \hat{n}_1 \cdot \vec{v}_2$, $pm_3 \equiv \hat{n}_2 \cdot \vec{v}_3$,
and $pm^*_1 \equiv \hat{n}_{-2} \cdot \vec{v}^*_1$, $pm^*_2 \equiv \hat{n}_{-3} \cdot \vec{v}^*_2$, $pm^*_3 \equiv \hat{n}_{-1} \cdot \vec{v}^*_3$), the six optical-path noises $y^\mathrm{op}_{slr}$, and the six laser noises $C_i$ and $C_i^*$.
(Note that our definition of the $pm^*_r$ differs by a factor $-1$ from the definition used in Ref.\ \cite{lisasimwhitepaper}.)

The general expressions for $y^\mathrm{noise}_{slr}$ and $z^\mathrm{noise}_{slr}$ then become
\begin{widetext}
\begin{eqnarray}
y^\mathrm{noise}_{slr}(t) = 
\left\{
\begin{array}{c}
C^*_s\bigr(t - L_l(t)\bigl) - C_r(t) + y^\mathrm{op}_{slr}(t) - 2 pm_r(t)
\quad \mbox{if $l > 0$}, \\[6pt]
C_s\bigr(t - L_l(t)\bigl) - C^*_r(t) + y^\mathrm{op}_{slr}(t) - 2 pm^*_r(t)
\quad \mbox{if $l < 0$},
\end{array}
\right.
\label{eq:ynoises}
\end{eqnarray}
and
\begin{eqnarray}
z^\mathrm{noise}_{slr}(t) = 
\left\{
\begin{array}{c}
C^*_r(t) - C_r(t) + 2 pm^*_r(t)
\quad \mbox{if $l > 0$}, \\[6pt]
C_r(t) - C^*_r(t) + 2 pm_r(t)
\quad \mbox{if $l < 0$}.
\end{array}
\right.
\label{eq:znoises}
\end{eqnarray}
\end{widetext}
We set standard levels for the 18 fundamental noises according to the noise budget discussed in the LISA pre-phase A report \cite{PPA98}. Note however that \emph{Synthetic LISA} allows all these prescriptions to be overridden.

%
\mylistitem{Laser Frequency Noise.} We take each laser noise to be white, and to have a one-sided (square-root) spectral density of $30$ $\mathrm{Hz} / \sqrt{\mathrm{Hz}}$, which converts to a power spectrum of fractional frequency fluctuations by squaring and dividing by the square of the optical frequency $\simeq c / (1064 \, \mathrm{nm}) = 2.82 \times 10^{14}$ Hz; thus, $S_n^\mathrm{ls} = 1.1 \times 10^{-26} \, \mathrm{Hz}^{-1}$. We assume that the six laser noises are statistically independent (the lasers need not be locked).

\mylistitem{Proof-Mass Noise.} We take each proof mass to have white acceleration noise along the line of sight, with a one-sided (square-root) spectral density of $3 \times 10^{-15}$ m s$^{-2}$ Hz$^{-1/2}$, which converts to a power spectrum of fractional frequency fluctuations \cite{ETA2000,tdiwhitepaper} by using the derivative theorem for Fourier transforms, and dividing by $c^2$; thus, $S^\mathrm{pm}_n = (3 \times 10^{-15}$ m s$^{-2}$ Hz$^{-1/2})^2$/($4 {\pi}^2 f^2 c^2)$ = $2.5 \times {10^{-48}} [f/\mathrm{Hz}]^{-2}$ Hz$^{-1}$. We assume that the six proof-mass noises are statistically independent.

\mylistitem{Optical-Path Noise.} We combine shot noise and beam-pointing noise on each optical bench into aggregate optical-path noises; we take these to be white displacement noises, with a one-sided (square-root) spectral density of $20 \times 10^{-12}$ m Hz$^{-1/2}$, which converts to a power spectrum of fractional frequency fluctuations by using the derivative theorem for the Fourier transform, and dividing by $c^2$; thus, $S^\mathrm{op}_n = (20 \times 10^{-12}$ m Hz$^{-1/2})^2 \times (4 {\pi}^2 f^2) / c^2 = 1.8 \times {10^{-37}}$ [f/Hz]$^2$ Hz$^{-1}$.  If the length of the LISA arms is different from the nominal value of 16.6782 s, we scale the optical-path rms noise by $L_l/(16.6782 \, \mathrm{s})$ to account for the $1/L_l^2$ power loss along the arms.\footnote{The variance of shot noise is inversely proportional to the number of photons received, which is proportional to the power received. Since power scales as $1/L_l^2$, rms shot noise must scale as $L_l$. We assume that the remaining part of the aggregate optical-path noise scales in the same fashion.}
We assume that our aggregate optical-path noise enters the $y_{slr}$ and $z_{slr}$ observables in the same way as shot noise, and we further assume that the six optical-path noises are statistically independent.
%

\subsection{LISA TDI observables}
\label{sec:tdi}
\begin{figure}\begin{center}
\includegraphics[width=3.2in]{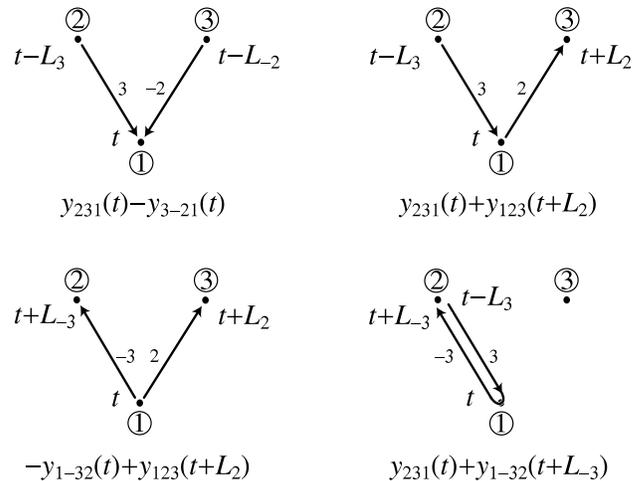}
\end{center}\caption{The four combinations of two basic Doppler observables with emission or reception at spacecraft 1 and at time $t$.\label{fig:tdivisual}}
\end{figure}

\emph{Time Delay Interferometry} \cite{TA1999,AET1999,ETA2000,TAE2000,AET2001,TEA2002,PTLA2002,algebraic2002,tdiwhitepaper,AET2003,TSSA2003,flexy2003,STEA2003,Shaddock2004,TEA2004} is a technique to combine the basic Doppler variables $y_{slr}$ and $z_{slr}$ into composite observables that are sensitive to GWs, but that are free of the otherwise overwhelming laser frequency noise (they are also free of optical-bench and fiber noise, as discussed above). To understand how TDI works, it is useful to tie the algebraic representation of the TDI observables to a visual picture of the path traveled by light between the LISA spacecraft. Looking at Fig.\ \ref{fig:tdivisual}, let us consider the four combinations $y_{231}(t) - y_{3-21}(t)$, $y_{231}(t) + y_{123}(t+L_2)$, $-y_{1-32}(t+L_{-3}) + y_{123}(t+L_{2})$, and $y_{231}(t)+y_{1-32}(t+L_{-3})$; the two laser beams involved in each of these combinations are either arriving to spacecraft $1$ or leaving it at the time $t$; the retardations by the armlengths $L_2$ and $L_{-3}$ are needed because the Doppler observables are always labeled by the time of beam \emph{reception}. Using Eq.\ \eqref{eq:ynoises}, we see that for the first three combinations the contribution to laser frequency noise that is produced \emph{at time $t$} is due to lasers 1 and $1^*$, and it is equal to $-C_1(t) + C^*_1(t)$. In the fourth combination, $y_{231}(t)+y_{1-32}(t+L_{-3})$, no laser noise is produced at time $t$, because the same laser is used as emitter and reference. For the first three combinations, the laser-noise contribution \emph{can be canceled} by subtracting from the $y_{slr}$ expressions given above the intra-spacecraft measurement $(1/2)[z_{231}(t) - z_{3-21}(t)]$, whose laser-noise component is again $-C_1(t) + C^*_1(t)$ (in fact, as noted above, each of $z_{231}(t)$ and $-z_{3-21}(t)$ contains the combination $-C_1(t) + C^*_1(t)$, but the difference of the two $z$ has the added advantage of canceling optical-fiber noise).

\begin{figure}\begin{center}
\includegraphics[width=3in]{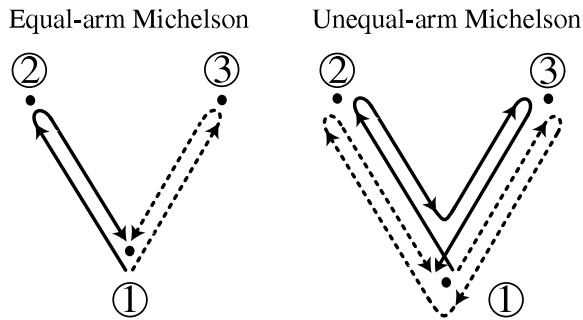}
\end{center}\caption{Tracing the light paths in the Michelson and unequal-arm Michelson TDI combinations.\label{fig:michelson}}
\end{figure}

Naturally, the laser noise that is produced at the times $t - L_3$, $t - L_{-2}$, $t + L_2$, and $t + L_{-3}$ (in various combinations for the four $y_{slr}$ expressions) is still not canceled. We see, however, that a combination of $y_{slr}$ observables that corresponds graphically \emph{to a closed circuit} would cancel laser noise completely; to build such a combination, we need to delay the times of evaluation for the $y_{slr}$ so that the tip or tail of each arrow meets another tip or tail (and only one!) at just the right time. A simple example, valid in the case when the $L_l$ are time-independent and all equal to $L$, traces a light path analogous to the path used in a Michelson interferometer (see the left panel of Fig.\ \ref{fig:michelson}),
\begin{widetext}
\begin{multline}
\bigl[ y_{123}(t + L_{2}) + y_{3-21}(t + L_{2} + L_{-2}) \bigr]
-\bigl[ y_{1-32}(t + L_{-3}) + y_{231}(t + L_{-3} + L_{3}) \bigr] \\
\quad - \onehalf \bigl(z_{231}(t) - z_{3-21}(t)\bigr) + \onehalf \bigl(z_{231}(t + L_{-3} + L_{3}) - z_{3-21}(t + L_{2} + L_{-2})\bigr),
\label{eq:simplemichelson}
\end{multline}
\end{widetext}
where the two interfering light beams leave spacecraft 1 at time $t$, and return at time $t + 2 L$. The two double-$z_{slr}$ subtraction terms are needed for the initial time of emission of the two beams, and for the final time of arrival, while laser noise is self-canceling at the zero-angle corners where the beams retrace their path, as mentioned above. Reordering Eq.\ \eqref{eq:simplemichelson} so that $t$ is the final time of arrival of the beams at spacecraft 1, we get
\begin{multline}
\bigl[ y_{3-21}(t) + y_{123,-2}(t) \bigr] - \bigl[ y_{231}(t) + y_{1-32,3}(t) \bigr] \\
- \onehalf \bigl(z_{231,-33}(t) - z_{3-21,2-2}(t)\bigr) + \onehalf \bigl(z_{231}(t) - z_{3-21}(t)\bigr),
\end{multline}
where the comma notation $y_{slr,d_1d_2\ldots}$ denotes retardation by the armlengths $L_{d_1}$, $L_{d_2}$, and so on. Laser-noise cancellation works in this case because the length of the two paths $1 \rightarrow 3 \rightarrow 1$ and $1 \rightarrow 2 \rightarrow 1$ is the same ($2 L$), so we can line up both the starting and the ending points of the two paths. If the arms (and hence the paths) were unequal, we would be left with residual laser noise originating from the starting points of the two paths, as given by
\begin{equation}
-\bigl(\onehalf C_{1,-33}(t) - \onehalf C_{1,2-2}(t)\bigr) +
\bigl(\onehalf C^*_{1,2-2}(t)  - \onehalf C^*_{1,-33}(t)\bigr), 
\end{equation}
The case of unequal (but constant) arms is tackled successfully by using new paths ($1 \rightarrow 3 \rightarrow 1 \rightarrow 2 \rightarrow 1$ and $1 \rightarrow 2 \rightarrow 1 \rightarrow 3 \rightarrow 1$) each of which traces out \emph{both} original paths ($1 \rightarrow 2 \rightarrow 1$ and $1 \rightarrow 3 \rightarrow 1$), but in opposite orders (see right panel of Fig.\ \ref{fig:michelson}). In this case, if we set the two paths to end at time $t$, the times of departure are both $t - (L_2 + L_{-2}) - (L_{-3} + L_{3})$, and the $z_{slr}$ correction terms can cancel the noise emitted at that time, as well as time $t$. The corresponding TDI combination (known as unequal-arm--Michelson $X$, and first derived by Tinto and Armstrong \cite{TA1999}) is
\begin{widetext}
\begin{eqnarray}
X & = & \bigl[ y_{1-32,32-2} + y_{231,2-2} + y_{123,-2} + y_{3-21} \bigr]
- \bigl[y_{123,-2-33} + y_{3-21,-33} + y_{1-32,3} + y_{231} \bigr]
\nonumber \\[1mm]
& & \quad - \onehalf\bigl(z_{3-21,2-2-33} - z_{231,-332-2}\bigr) - \onehalf\bigl(z_{3-21} - z_{231} \bigr)
\\[1mm]
& & \quad + \onehalf\bigl(z_{3-21,2-2} - z_{231,2-2}\bigr)
+ \onehalf\bigl(z_{3-21,-33} - z_{231,-33}\bigr)
\nonumber \label{eq:X}
\end{eqnarray}
\end{widetext}
where we omitted the dependence on $t$ common to all the terms.

Many TDI combinations are possible: all cancel laser noise, but each shows a different coupling to GWs and to the remaining system noises (known collectively as \emph{secondary noises}). As the understanding of TDI improved, the standard TDI observables evolved through various generations, capable of canceling laser noise for increasingly complex LISA geometries:

%
\mylistitem{First-generation TDI.} Also known as TDI 1.0. The first-generation TDI observables \cite{TA1999,AET1999,ETA2000,AET2001} cancel laser noise exactly in LISA configurations with unequal (but constant) arms, and $L_k = L_{-k}$. Interferometric combinations of various types are possible:

The \emph{Sagnac-type observables} ($\alpha$, $\beta$, $\gamma$) are sums of six basic Doppler observables, and they involve the difference between the Doppler shifts accumulated by light propagating around the LISA array in the two senses. Thus, the Sagnac-type observables use all the LISA laser links in both directions. A fully symmetric Sagnac observable ($\zeta$) is considerably less sensitive than most others to GWs with frequencies at the lower end of the LISA band; it was suggested \cite{TAE2000} that the comparison between the power observed in $\zeta$ and in the other TDI variables could be used to discern a stochastic GW background from instrumental noise. The observables built from six Doppler variables are also known as \emph{six-pulse combinations}, because their response to an impulsive plane GW consists of six separate pulses.

\emph{Eight-pulse combinations} involve sums and differences of the Doppler shifts measured along four of the six LISA laser links. The \emph{unequal-arm Michelson observables} ($X$, $Y$, $Z$) use both links of two arms; as discussed above, they can be interpreted as measuring the phase difference accumulated by light traveling (twice, in opposite orders) along the two arms of a Michelson interferometer centered in one of the spacecraft. Perhaps for this reason, and in analogy with ground-based GW interferometers, a single unequal-arm Michelson observable (generally $X$) is often used in LISA data analysis to compute expected detection rates and parameter-estimation accuracies.

\emph{More eight-pulse combinations} can be formed: the \emph{beacon} observables ($P$, $Q$, $R$) use only the two links departing from one of the spacecraft, and both links along the opposite arm; the \emph{monitor} observables ($E$, $F$, $G$) use only the two links arriving at one of the spacecraft, and both links along the opposite arm; last, the \emph{relay} observables ($U$, $V$, $W$) use one departing link and the adjacent arriving link at one of the spacecraft, together with both links along the opposite arm. The eight-pulse combinations can be considered as LISA contingency modes, because they are available even if one or two of the laser links fail. Note however that all six lasers must still be available to build the intra-spacecraft observables $z_{slr}$ required for the eight-pulse combinations, except in the case of the unequal-arm Michelson observables: one of these can always be built even if one or both lasers directed along one of the arms happen to fail.

Dhurandhar and colleagues \cite{algebraic2002} proved that the space of all the first-generation TDI observables can be constructed by combining four generators, which they identify in $\alpha$, $\beta$, $\gamma$, and $\zeta$. Prince and colleagues \cite{PTLA2002} showed how to diagonalize the cross noise spectrum of the generators to obtain three observables ($A$, $E$, and $T$) \emph{with uncorrelated noises}. The three \emph{optimal observables} $A$, $E$, and $T$ are written as sums and differences of $\alpha$, $\beta$, and $\gamma$, and when used in combination they achieve the optimal S/N for GW sources at any frequency in the LISA band.

\mylistitem{Modified TDI.} Also known as TDI 1.5. Shaddock \cite{Shaddock2004} recently pointed out that the rotation of the LISA array introduces a difference in the armlengths experienced by beams traveling in the corotating and counterrotating directions (i.e., $L_k \neq L_{-k}$). Furthermore, this difference becomes much larger if we take into account also the orbital motion of the array around the Sun \cite{TEA2004}. Some of the first-generation observables (the $X$-type, $P$-type, $E$-type, and $U$-type combinations), cancel laser noise also for $L_k \neq L_{-k}$, if time delays for the appropriate oriented arms are used [as we have already arranged, for instance, in Eq.\ \eqref{eq:X}]; these observables can be interpreted as tracing light paths that enclose vanishing areas. Conversely, the first-generation observables that trace light paths that enclose a finite area (such as $\alpha$, $\beta$, $\gamma$, and $\zeta$) are equivalent to \emph{Sagnac interferometers} \cite{sagnac}, and must necessarily be sensitive to the rotation of the array, which shows up as a spurious phase difference between the lasers, originating from the starting points of the light paths. The Sagnac observables can be modified by means of a finite-difference procedure analogous to the change undergone between the equal-arm and unequal-arm Michelson combinations (see Fig.\ \ref{fig:michelson}), so that the modified Sagnac observables have null enclosed area, and cancel laser noise \cite{flexy2003,STEA2003}. The resulting combinations [$\alpha_1$, $\alpha_2$, and $\alpha_3$, which generalize $\alpha$, $\beta$, and $\gamma$; and $\zeta_1$, $\zeta_2$, and $\zeta_3$ \cite{TEA2004}, which nonuniquely generalize $\zeta$] include twice as many $y_{slr}$ variables as the first-generation combinations (i.e., they are 12-pulse observables).

\mylistitem{Second-generation TDI.} Also known as TDI 2.0. The motion of the LISA array introduces not only a directional dependence of the armlengths, but also a time dependence, as first recognized by Cornish and Hellings \cite{flexy2003}. In this case, the order of the TDI retardations becomes important: for instance, if the armlengths are constant, then
\begin{equation}
t_{,2-2} \equiv t - L_{-2} - L_{2} \; =
\; t - L_{2} - L_{-2} \equiv t_{,-22}
\end{equation}
but if they are not (as signaled by a semicolon index notation), then
\begin{multline}
t_{;2-2} \equiv \bigl(t - L_{-2}(t) - L_{2}(t - L_{-2})\bigr)
\\ \neq \bigl(t - L_{2}(t) - L_{-2}(t - L_{2})\bigr) \equiv t_{;-22}.
\end{multline}
More generally, the semicolon notation represents the retardation chain rule
\begin{multline}
\label{eq:chainrule}
t_{;d_1\ldots d_n} = t - L_{d_n    }(t) - L_{d_{n-1}}\big(t - L_{d_n    }(t)\big)
\\ - L_{d_{n-2}}\Big(t - L_{d_n    }(t) - L_{d_{n-1}}\big(t - L_{d_n    }(t)\big)\Big) - \cdots  
\end{multline}
where the rightmost retardation index is applied first, using the armlength $L_{d_n}(t)$; the next-to-rightmost retardation index is applied second, using the partially retarded armlength $L_{d_{n-1}}(t-L_{d_n}(t))$, and so on. Taylor-expanding the armlengths, and retaining only the zeroth-order and first-order terms, we get
\begin{multline}
t_{;d_1\ldots d_n} = t - L_{d_n}
- \left[ L_{d_{n-1}}
- \dot{L}_{d_{n-1}} L_{d_n} \right]
\\ - \left[ L_{d_{n-2}}
- \dot{L}_{d_{n-2}}\left( L_{d_n} + L_{d_{n-1}} \right) \right]
- \cdots
\end{multline}
where for ease of notation we have dropped the $(t)$ dependence common to all the armlengths. As discussed in Refs.\ \cite{flexy2003,STEA2003,TEA2004}, the eight-pulse TDI observables can be generalized, once again by a procedure akin to finite differentiation, to 16-pulse observables that cancel laser noise up to first order in the Taylor-expanded armlengths; for the LISA orbital parameters, this is enough to cancel laser noise to a level below the secondary noises. According to the notation of Ref.\ \cite{TEA2004}, $X_1$, $X_2$, and $X_3$ generalize $X$, $Y$, and $Z$; $P_1$, $P_2$, and $P_3$ generalize $P$, $Q$, and $R$; $E_1$, $E_2$, and $E_3$ generalize $E$, $F$, and $G$; and $U_1$, $U_2$, and $U_3$ generalize $U$, $V$, and $W$. The $X_k$ observables can be interpreted as expressing the difference in laser phase between beams propagating along two paths whose Taylor-expanded total lengths differ only by terms proportional to $\ddot{L}_k$ or to higher derivatives;\footnote{The finite differencing procedure adopts the compound paths A $\equiv$ I+II and B $\equiv$ II+I, where the paths I and II must contain the same links, in different orders; then $t_{\mathrm{I,II}} - t_{\mathrm{II,I}} \simeq \dot{\mathrm{I}}\times\mathrm{II} - \dot{\mathrm{II}}\times\mathrm{I} \equiv (\sum_i \dot{L}_{\mathrm{I}_i})(\sum_j L_{\mathrm{II}_i}) - (\sum_i \dot{L}_{\mathrm{II}_i})(\sum_j L_{\mathrm{I}_i}) = 0$.} the residual laser noise is then a sum of expressions similar to
\begin{multline}
C_{k;\mathrm{A}}(t) - C_{k;\mathrm{B}}(t)
\;\simeq\;
\dot{C}_{k}(t) \times \left[ t_{;\mathrm{A}} - t_{;\mathrm{B}} \right]
\\ \simeq\;
\dot{C}_{k}(t) \times O\bigl[\ddot{L} \; \mbox{and higher derivatives}\bigr].
\end{multline}
As for the Sagnac-type observables, the 12-pulse modified observables $\alpha_1$, $\alpha_2$, $\alpha_3$, $\zeta_1$, $\zeta_2$, and $\zeta_3$ can already cancel laser noise to a level below the LISA secondary noises: the residual laser noise is of order $\dot{L}$ and higher, but the specific combination of $\dot{L}_k$ involved turns out to be small for the LISA orbit.
%

Although historically the TDI observables were derived by combining time-shifted combinations of the basic (one-way) Doppler measurements $y_{slr}(t)$, they can also be written as combinations of one-way and two-way Doppler measurements, generated by locking five of the six lasers to the remaining one, as described by Tinto and colleagues \cite{TSSA2003}; the resulting expressions contain fewer terms, are still noise-canceling, and have the same response to GWs.

\section{Implementation and usage of \textit{Synthetic LISA}}
\label{sec:synthlisa}

\textit{Synthetic LISA} is an object-oriented C++ library built to
mirror the idealized structure of Fig.\ \ref{fig:lisablock}: each block in the figure corresponds to one or more C++ classes \cite{cplusplus}, which implement its functionality. The \textit{Synthetic LISA} workflow follows this object-oriented structure, facilitating targeted investigations that compare multiple configurations of one object (for instance, one of the fundamental noises, or the GW source), while all others are kept fixed. Here is an example of a typical \textit{Synthetic LISA} session.
\begin{enumerate}
\item \emph{Create an instance of a LISA geometry (\texttt{LISA}) class with the desired orbital parameters.}

The \texttt{LISA} classes provide the geometrical quantities $\vect{p}_i(t)$, $\hat{n}_l(t)$, and $L_l(t)$ needed to assemble the LISA GW and noise responses described in Secs.\ \ref{sec:gwresponse} and \ref{sec:noiseresponse}. They account for the aberration effects caused by the finite speed of light and by the spacecraft motion intervening between the events of pulse emission and reception [Eq.\ \eqref{eq:lightprop}].

There are different levels of complexity at which the motion of the LISA array, discussed in Sec.\ \ref{sec:lisamodel}, can be modeled in a simulation of the LISA science process; correspondingly, increasingly sophisticated TDI observables are needed to cancel laser noise once the added complexity is taken into account. In \emph{Synthetic LISA}, these levels correspond to different derived classes\footnote{In C++, a derived class inherits the data content and behavior of its base class, and can add enhancements or customizations.} of the base class \texttt{LISA}. The simplest such class, \texttt{OriginalLISA}, models a stationary, nonorbiting constellation, used implicitly in the development of first-generation TDI. The most realistic, \texttt{EccentricInclined}, models the eccentric orbits of the spacecraft up to second order in the eccentricity (see App.\ \ref{app:conventions}); the resulting time dependence of the armlengths \cite{flexy2003} creates the necessity of second-generation TDI for effective laser-noise suppression.
\item \emph{Create instances of a LISA noise class (\texttt{Noise}) for the 18 fundamental-noise time series defined in Sec.\ \ref{sec:noiseresponse}, tuning noise parameters if so desired.}

\textit{SyntheticLISA} can generate pseudorandom noise sequences that approximate closely the standard laser, proof-mass, and optical-path noises specified in Sec.\ \ref{sec:noiseresponse}; alternatively, the package can import the noises as sampled time series, which might have been generated with other tools, or even measured experimentally. The treatment of the LISA noise processes is crucial to the interpretation of \emph{Synthetic LISA} simulations, and is discussed in detail in App.\ \ref{app:synthnoise}. In short, the representation of noise is adequate if the noise-generation Nyquist frequency $f_b$ is set comfortably higher than the highest frequency at which one wishes to analyze the TDI noise responses, but of course lower than the Nyquist frequency used to sample the TDI observables, to avoid aliasing.
\item \emph{Create an instance of a GW source class (\texttt{Wave}) of the desired type and parameters.}

The \texttt{Wave} classes provide the GW polarization components $h_+(t)$ and $h_\times(t)$, which are assembled into the transverse--traceless metric perturbation $\mathsf{h}(t)$ according to the polarization convention described in App.\ \ref{app:conventions}. \textit{Synthetic LISA} contains simple \texttt{Wave} classes (such as \texttt{SimpleBinary} for monochromatic binaries), which can be modified easily to yield more complicated signals; the package can also import $h_\times(t)$ and $h_\times(t)$ as sampled time series.
\item \emph{Create an instance of a LISA TDI class (\texttt{TDI}), feeding it the LISA geometry, LISA noises, and GW source objects previously created.}  

The base class \cpp{TDI} defines a complete set of first-generation, modified, and second-generation TDI observables, according to the expressions of Refs.\ \cite{AET1999,ETA2000} for first-generation TDI, and of Refs.\ \cite{TEA2004,STEA2003} for modified and second-generation TDI.\footnote{
The primed link indices of Refs.\ \cite{TEA2004,STEA2003} correspond to positive indices in this paper.} The derived classes \texttt{TDIsignal} and \texttt{TDInoise} implement, respectively, the LISA response to GWs [$y^\mathrm{gw}_{slr}$ from Eq.\ \eqref{eq:responses}], and to the fundamental noises [$y^\mathrm{noise}_{slr}$ and $z^\mathrm{noise}_{slr}$ from Eqs.\ \eqref{eq:ynoises} and \eqref{eq:znoises}]. Users can easily define additional TDI observables, using Table \ref{tab:rosetta} to rewrite the expressions in the literature in terms of the $y_{slr}$ and $z_{slr}$ \emph{Synthetic LISA} observables.
\item \emph{Last, use the TDI objects to generate a time series of the TDI observables and write it to disk or memory.}
\end{enumerate}
No C++ programming and compilation is needed to use \emph{Synthetic LISA}, since the functionality of the package can be accessed very easily from the scripting language Python \cite{python}, either interactively, or with short scripts. In fact, the \textit{Synthetic LISA} session described above would translate to a handful of lines in Python. Refer to the \emph{Synthetic LISA} manual \cite{synthlisamanual} for detailed information about the usage and implementation of the package. The manual documents also the successful validation of \emph{Synthetic LISA}'s output against analytical expressions of the TDI observables for both noise and signals.

\section{Numerical experiments with \textit{Synthetic LISA}}
\label{sec:tests}

We now present the main scientific results of this paper: an investigation of laser phase noise suppression for flexing LISA array orbits with first- and second-generation TDI [Sec.\ \ref{sec:expsecondgen}]; an analysis of the armlength-determination accuracies required for effective laser-noise suppression [Sec.\ \ref{sec:exparmlengths}]; and an estimation of quantization and telemetry bitdepth 
needed for the phase measurements $y_{slr}$ and $z_{slr}$. While significant as they stand, these studies are meant also to exemplify the kind of system-characterization inquiries that becomes possible with advanced LISA simulators.

Except where otherwise specified, all the power spectra displayed in this section were computed as periodograms, reducing spectral leakage and fluctuations by dividing one-year--long time series into partially overlapping segments (in number of either 1024 or 2048, depending on the specific test), triangle-windowing each segment, and averaging the resulting power spectra (see, e.g., Ref.\ \cite{nrc}). Thus, all the spectra of this section represent \emph{average} effects: slightly different requirements on laser-noise power, armlength determination, and phase-measurement quantization might be needed to achieve the same suppression performance homogeneously across the year.

\subsection{On the necessity of second-generation TDI}
\label{sec:expsecondgen}

As recognized by Cornish and Hellings \cite{flexy2003}, the eccentric and inclined orbital motion of the LISA spacecraft introduces a time variation in the armlengths of order $10^{-8}$ s/s [see Eq.\ \eqref{eq:flexyarms} of App.\ \ref{app:conventions}]; as a consequence, the first-generation and modified TDI observables fail to cancel laser frequency noise completely. Using the graphical interpretation of TDI given in Sec.\ \ref{sec:tdi}, we would see that the interferometric circuits synthesized by the observables fail to close exactly. The laser-noise residuals arise from the starting points of the paths, and they are of the form
\begin{eqnarray}
\label{eq:lnres}
\delta C_i &=& \onehalf [C^*_{i;J}(t) - C^*_{i;I}(t)]
- \onehalf [C_{i;I}(t) - C_{i;J}(t)] \\
&\simeq& \onehalf [\dot{C}^*_i(t) + \dot{C}_i(t) ] (t_{;J}-t_{;I}) =
\onehalf [\dot{C}^*_i(t) + \dot{C}_i(t) ] \delta t, \nonumber
\end{eqnarray}
where $I$ and $J$ denote time-ordered path retardation chains.
Using the Fourier derivative theorem and assuming white,
uncorrelated laser noises, we get
\begin{equation}
\label{eq:lnress}
|\delta \tilde{C}_i(f)|^2/|\tilde{C}_i|^2 = 2 \pi^2 f^2 \delta t^2,
\end{equation}
for frequencies up to the laser-noise bandlimit. For the
modified TDI X observable, $\delta t \simeq 10^{-6}$ s, so laser noise is canceled by less than 160 dB for $f \gtrsim 2$ mHz; for the second-generation TDI $X_1$ observable, $\delta t \simeq 10^{-10}$ s, so laser noise is canceled comfortably by more than 160 dB throughout the LISA band of good sensitivity. In this section we discuss the results of \emph{Synthetic LISA} simulations carried out to investigate and substantiate these analytic arguments.

\begin{figure}\begin{center}
\includegraphics[width=3.4in]{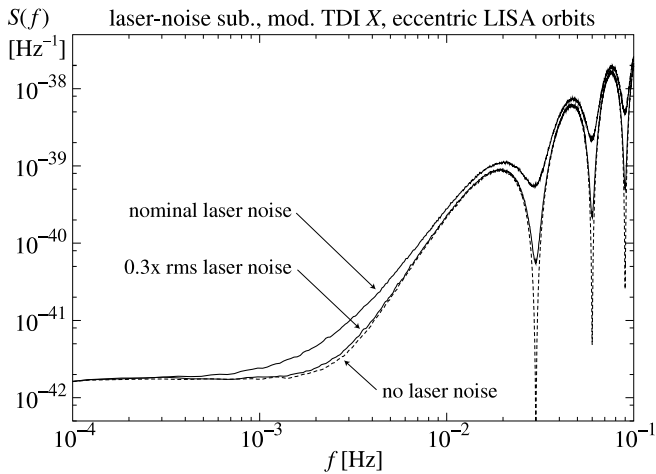}
\end{center}\caption{
Imperfect cancellation of noise with modified TDI
$X$ for flexing LISA array (\texttt{EccentricInclined}, $\xi_0 = \eta_0 = 0$). The top curve plots the perfect-cancellation noise target, obtained by setting the laser noise to zero; the intermediate and bottom curves show the noise spectra resulting from imperfect laser-noise cancellation for nominal and $0.3 \times$ rms laser noise. A  curve with $0.1 \times$ rms laser noise would be essentially indistinguishable from the perfect-cancellation target, and is not plotted here. The spectra are computed from one year's worth of $X$ data sampled at 1 Hz (with 1-s noise-generation timestep), averaging over 2048 data subsegments.\label{fig:tdi2ndX}}
\end{figure}

Figure \ref{fig:tdi2ndX} shows the spectrum of secondary noise plus residual laser noise (top curve) versus the spectrum of secondary noise alone for the modified TDI $X$ observable (bottom curve), computed using realistic eccentric and inclined LISA spacecraft orbits; the excess noise is evident between 1 mHz and 10 mHz, and within the noise nulls at the frequency multiples of $1/(2L)$. The intermediate curve shows the effect of reducing the laser noise to 0.3 times its nominal rms spectral density $1.1 \times 10^{-26} \, \mathrm{Hz}^{-1}$. A separate simulation was performed by reducing laser noise to 0.1 times its nominal value: the resulting spectrum is essentially indistinguishable from the secondary-noise--only curve.

The reader might be puzzled by the flatness of the first-generation TDI noise curves at low frequencies, as compared to the $f^{-2}$ dependence of proof-mass noise and of the often-seen LISA \emph{sensitivity} curves. The flatness is caused by the time-delay structure of first-generation TDI observables, which contain, as it were, a finite-difference time derivative, with a low-frequency power transfer function proportional to $f^2$. On the other hand, the sensitivity curves plot a ratio of (rms) noise response to GW-signal response, with the latter decreasing as $f^2$ at low frequency for first-generation TDI observables such as $X$ \cite{ETA2000}.

\begin{figure}\begin{center}
\includegraphics[width=3.4in]{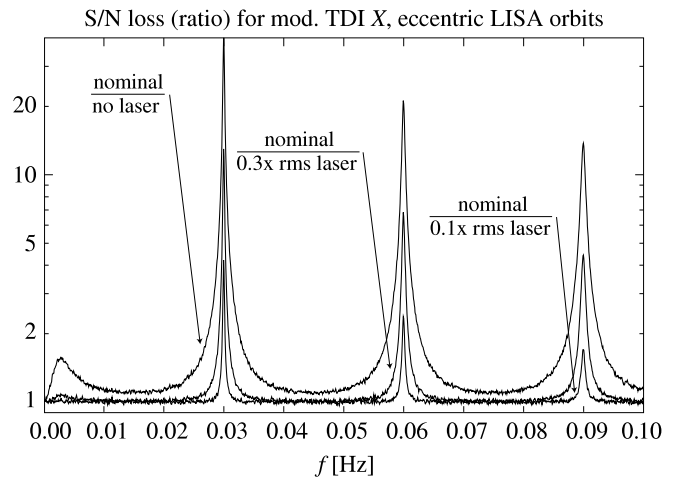}
\end{center}\caption{
Reduction in amplitude S/N due to imperfect lasernoise
cancellation for modified TDI $X$, with realistic LISA
spacecraft orbits (\texttt{EccentricInclined}, $\xi_0 = \eta_0 = 0$) and with nominal laser noise, $0.3 \times$ rms laser noise, and $0.1 \times$ rms laser noise. Because of the linear frequency axis used in this graph, the bump between 1
mHz and 10 mHz (which tops at about 1.3) appears unimpressive,
but it spans a scientifically important frequency range. (From the same data as Fig.\ \ref{fig:tdi2ndX}.)\label{fig:test-tdi2ndSN}}
\end{figure}

Figure \ref{fig:test-tdi2ndSN} shows the reduction caused by
residual laser noise in the (amplitude) S/N for monochromatic
sources, computed as the square-root ratio of the imperfect-cancellation and secondary-noise--only spectra. The loss of sensitivity
appears significant (up to $\sim 30\%$) between 1 mHz and
10 mHz, and even more so around the $1/(2L)$ harmonics. However,
an improvement in laser noise stability by a factor
of about three would be sufficient to erase the S/N-reduction
bump at lower frequencies, and to shrink considerably
the S/N-reduction peaks at higher frequencies.
An improvement by a factor of ten would essentially eliminate
the need for second-generation TDI, as estimated
analytically in Ref.\ \cite{flexy2003}.

\begin{figure}\begin{center}
\includegraphics[width=3.4in]{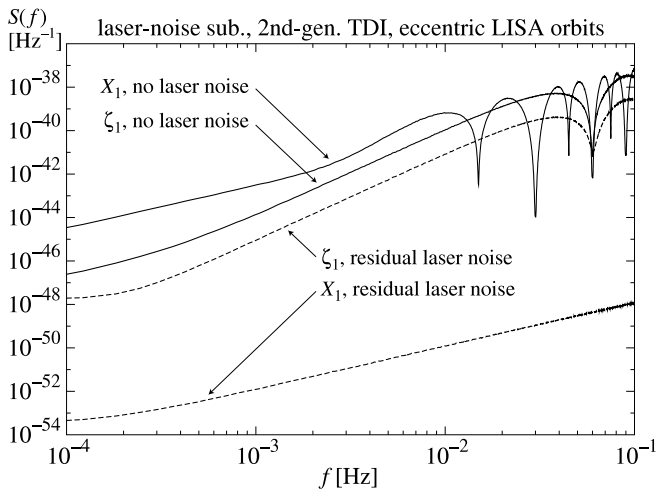}
\end{center}\caption{
Cancellation of laser noise with second-generation TDI
$X_1$ and modified TDI $\zeta_1$ for flexing LISA array (\texttt{EccentricInclined}, $\xi_0 = \eta_0 = 0$). For $X_1$, the noise spectrum obtained by including nominal laser noise matches exactly the perfect-cancellation noise target obtained by setting the laser noise to zero (top curve); the bottom curve, several orders of magnitude below, shows the spectrum of residual laser noise.
For $\zeta_1$, residual laser noise sits only one order of magnitude below the perfect-cancellation noise target.
The spectra are computed from one year's worth of $X_1$ and $\zeta_1$ data sampled at 1 Hz (with 1-s noise-generation timestep), averaging over 2048 data subsegments.\label{fig:test-tdi2ndSNX1}}
\end{figure}

By contrast, Fig.\ \ref{fig:test-tdi2ndSNX1} shows that essentially perfect laser-noise cancellation is achieved with the second-generation TDI observable $X_1$, with residual laser noise (bottom curve) several orders of magnitude below the secondary noises. For the Sagnac observable $\zeta_1$ (which, strictly speaking, belongs to the set of modified TDI observables), laser noise is still canceled by more than one order of magnitude, except at the first $\zeta_1$ null near $6 \times 10^{-2}$ Hz.

\subsection{On the required accuracy of armlength determination}
\label{sec:exparmlengths}
\begin{figure*}\begin{center}
\includegraphics[width=7in]{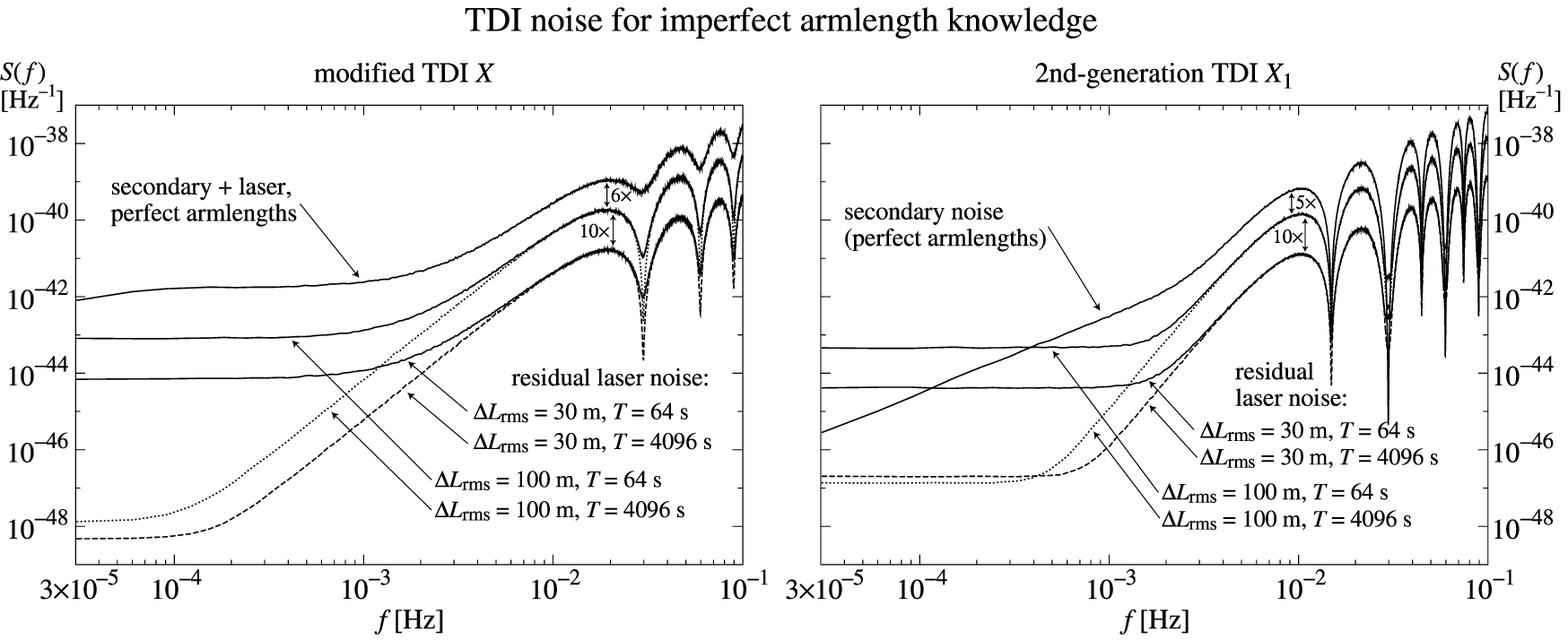}
\end{center}\caption{
Imperfect cancellation of laser noise with modified TDI $X$ (left panel) and second-generation TDI $X_1$ (right panel) due to imperfect knowledge of the armlengths in a flexing LISA array (\texttt{EccentricInclined}, $\xi_0 = \eta_0 = 0$).  
The topmost curves show the result of using perfect armlengths: thus, the $X_1$ curve shows secondary noise only, while the $X$ curve shows secondary noise plus the residual laser noise due to using modified TDI observables with a flexing LISA array. All other curves show the level of residual laser noise for linearly-extrapolated armlengths (see main text) with different single-measurement errors $\Delta L_\mathrm{rms}$ and intervals $T$. Error-laden armlength measurements are simulated by adding a Gaussian-distributed, zero-mean independent deviates to the correct values of the six $L_i$.  The low-frequency flattening of the laser-noise residuals is caused by power leakage from high frequencies when the bandwidth of the armlength-error time series is comparable with the LISA measurement bandwidth. The spectra are computed from one year's worth of $X$ and $X_1$ data sampled at 2 Hz (with 0.5-s noise-generation timestep), averaging over 1024 data subsegments.
\label{fig:test-noisyarms}}
\end{figure*}

We now turn to simulating the laser-noise residuals resulting from the implementation of TDI using an imperfect knowledge of the armlengths, and therefore of the TDI delays. In this case, residuals are created at all the delayed times that appear in the TDI observables, and not just at the starting point of the interferometric circuits. Graphically, the reason is that the tail of each $y_{slr}$ arrow (determined by a \emph{physical} light-travel delay) does not precisely match the head of the preceding arrow (determined by a \emph{nominal} TDI delay affected by armlength-measurement error). At each such point, the residual has the form
\begin{equation}
\label{eq:noisyresidue}
\delta C_s \simeq \dot{C}_s(t) \delta L_l(t).
\end{equation}
The total residual noise is a somewhat complicated function of the TDI observable under consideration.

Tinto and colleagues \cite{TSSA2003} find that an armlength accuracy of $\sim 30$ m ($\sim 100$ ns) would be needed for effective laser-noise subtraction with first-generation TDI.  They also estimate how often the armlength measurements would have to be updated, by computing the timescale for the time-dependent armlengths to change by an amount equal to the required accuracy; for realistic LISA orbits, this timescale varies substantially through the year, but it can be as low as $10$ s.

In the course of the LISA mission, armlengths might be determined by means of orbital-dynamics models that are periodically updated by ranging measurements, either between the spacecraft, or to Earth. It was recently suggested \cite{lisainterpolation} that the TDI observables do not need to be assembled in \emph{real time} aboard the spacecraft, but that they can be obtained in \emph{postprocessing} from time series of the $y_{slr}$ and $z_{slr}$ measurements sampled at limited rates ($\sim$ 1 Hz) and telemetered to Earth. If that is the case, the accuracy of ranging is probably a secondary issue, since even poor measurements can be fitted \emph{a posteriori} to very accurate models of the LISA orbits. In fact, it was recently proposed \cite{tdir} that the ranging information can be obtained directly from the $y_{slr}$ and $z_{slr}$ measurements, by minimizing the integrated noise power in the TDI observables as a function of the orbital parameters of the LISA spacecraft. Because of these reasons, the problem of determining the accuracy required for ranging measurements is not well defined in the context of postprocessed TDI. In this section we concentrate instead on the accuracy required for the real-time onboard computation of the TDI observables.

The simplest real-time treatment of the TDI delays consists simply of keeping the armlengths fixed to their last measured values, which are updated at time intervals $T$. The resulting requirements on the ranging measurements are rather constraining: for modified TDI $X$, our simulations show that marginally acceptable laser-noise cancellation is obtained with measurements repeated every 8 s with 2-m (rms) accuracies (assuming independent errors). Indeed, the piecewise-constant armlength model does a very poor job of following the dominant linear time dependence of the armlengths.
\begin{figure}\begin{center}
\includegraphics[width=3.4in]{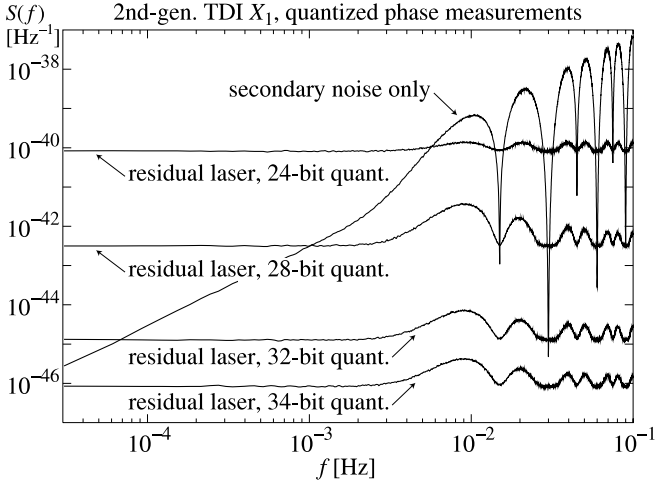}
\end{center}\caption{
Imperfect cancellation of laser noise with second-generation TDI $X_1$ due to quantization of the phase measurements using $n_\mathrm{quant}$ bits (see main text). The strongly sloping curve plots only the secondary noises, while all other curves show the level of residual laser noise with different quantization depths (the numbers shown do not include the additional 1 + 3 bits needed for the sign and to avoid saturation). The spectra are computed from one year's worth of $X$ and $X_1$ data sampled at 2 Hz (with 0.5-s noise-generation timestep), averaging over 1024 data subsegments.
\label{fig:test-quantize}}
\end{figure}

A better treatment, which requires very little sophistication in the onboard logic, consists of extrapolating linearly from the latest two armlength measurements, which are again repeated at intervals $T$. The left panel of Fig.\ \ref{fig:test-noisyarms} shows that, for modified TDI $X$, 100-m (rms) accurate measurements, repeated only every 4096 s, yield residual laser-noise suppression to better than a factor of six below the case of perfect armlength knowledge (where some residual laser noise is present because of the LISA array flexing; see Fig.\ \ref{fig:tdi2ndX}). Every successive $n$-fold improvement in the accuracies yields an $n^2$-fold improvement in laser-noise suppression. 

\addtolength{\parskip}{-0.75mm}
Remarkably, taking ranging measurements more often has the effect of \emph{worsening} laser-noise suppression at low frequencies. To understand why, consider Eq.\ \eqref{eq:noisyresidue}, which implies that in the Fourier domain the laser-noise residual is given by the \emph{convolution} of the laser-noise derivative with the armlength error. The rapid repetition of measurements introduces high-frequency power in the armlength-error time series with a typical bandwidth of $1/(2T)$, which then causes the leakage of power from high frequencies (where $\dot{C}_s$ is much larger) to the low-frequency end of the LISA spectrum. This behavior can be observed in the left panel of Fig.\ \ref{fig:test-noisyarms} by comparing the laser-noise residual curves corresponding to measurements repeated every 4096 s and every 64 s. By contrast, the maximum acceptable spacing of the measurements is set by the timescale for relevant quadratic changes in the armlengths: for a typical armlength acceleration $a \sim (2 \pi / \mathrm{yr}) \times 10^{-8} \, \mathrm{s}/\mathrm{s} = 2 \times 10^{-15} \, \mathrm{s}^{-1}$, the time required to accrete an error $\sim$ 100 m is $\sim \sqrt{2 \times 100 \, \mathrm{m} / a} =$ 18,000 s.

The right panel of Fig.\ \ref{fig:test-noisyarms} shows that the armlength accuracy requirement for the second-generation TDI observable $X_1$ is not substantially different, with 100-s (rms) accuracy achieving laser-noise suppression by a (power) factor of about five, and successive $n$-fold accuracy improvements yielding $n^2$-fold suppression improvements. However, considerably better accuracy is needed if laser noise is to be canceled also within the $X_1$ nulls at $1/(2L)$ and multiple frequencies. The leakage effect discussed above is more important in the case of second-generation TDI, where at low frequencies secondary noise declines as a positive power of $f$, and can intersect the leakage plateau if measurements are not taken sparsely enough.

\subsection{On the quantization of phase measurements}
\label{sec:quantize}

Our last numerical experiment in this paper is concerned with estimating the number of effective bits that must be obtained and recorded for the phase measurements $y_{slr}$ and $z_{slr}$, and then either transmitted between the spacecraft to perform TDI in real time, or transmitted to Earth to perform it in post processing.
\addtolength{\parindent}{0.75mm}

Similar, less extensive experiments have been performed by J.\ W.\ Armstrong \cite{jwaquant}. The underlying physical problem is that laser noise must be represented faithfully enough to allow its cancellation by several orders of magnitude.  Thus, we expect the spectral characteristics of laser noise, such as its bandwidth at the output of the phasemeter, and its magnitude relative to the secondary noises, to play into the answer to our question.  Presumably, considerable telemetry bandwidth can be saved by \emph{whitening} phase noise prior to transmission, in such a way that the quantity of (Fourier-space) information relative to secondary noise is approximately constant at all frequencies. For the purpose of our estimates, we adopt the crude whitening scheme implicit in dealing directly with fractional-frequency fluctuations; in this paper we assume laser noise to be white for these.

We quantize phase measurements by dividing each $y_{slr}$ and $z_{slr}$ (before assembling the TDI observables) by a fiducial fractional-frequency--fluctuation level given by the nominal rms value of laser noise (i.e., $1.05 \times 10^{-13}$, assuming noise bandlimited at 1 Hz), truncating the resulting values to $n_\mathrm{quant}$ bits to the right of the binary point, and then multiplying again by the fiducial level. The actual counting of bits must include one additional sign bit, and a few bits to the left of the binary point (we take three, which is adequate to make the truncation of Gaussian-distributed noise statistically insignicant). Figure \ref{fig:test-quantize} shows the results of our simulations for second-generation TDI $X_1$: an $n_\mathrm{quant}$ between 32 and 34 (and hence a total number of bits between 36 and 38) is needed to lower the level of residual laser noise resulting from quantization to a level comfortably below the secondary noises in the LISA measurement band. The requirement is less strict ($n_\mathrm{quant}$ between 28 and 30) for modified TDI $X$.

More definitive simulations of the effects of measurement quantization should include less idealized models of phase noise at the output of the phasemeter. Note also that the simulations presented here do not address the interplay between quantization and the implementation of fractional-filtering interpolation, used in postprocessed TDI \cite{lisainterpolation} to approximate the values of $y_{slr}$ and $z_{slr}$ at the TDI delays between recorded samples.

\section{Conclusions}
\label{sec:conclusions}

We have described three numerical experiments on the implementation of TDI in LISA, which were performed with \emph{Synthetic LISA}, a simulation of the LISA science process that can generate synthetic time series of fundamental noises and GW signals, as they appear in the laser-noise-canceling TDI observables. Our conclusions were presented in brief in Sec.\ \ref{sec:intro}, and described in detail in Sec.\ \ref{sec:tests}. We have also discussed the theoretical model that underlies \emph{Synthetic LISA} and provided details of its implementation, as needed to understand the results of our numerical experiments.

The structure and programming style used for \emph{Synthetic LISA} allows for vast extensions and improvements. Among others, we plan to include explicitly the additional time series required for calibration of the onboard ultrastable oscillators \cite{TEA2002}, and to model explicitly the measurement errors at the photodetectors.  
We are in the process of making \emph{Synthetic LISA} available \cite{synthlisa} as a public-domain software package, to foster the involvement of the wider GW community in research on the interface between scientific goals and technical requirements for LISA, on the tradeoffs and improvements that can be made in the implementation and operation of the mission, and on the development of novel analysis techniques for the LISA data. In the spirit of open-source design, we expect the LISA and GW communities to provide their own useful additions to \emph{Synthetic LISA}, such as more realistic models of the noises and of the spacecraft subsystems, and additional GW source modules. For this purpose, we have designed \emph{Synthetic LISA} as a modular and easily extensible C++ package, with a user-friendly Python frontend for easy scripting and prototyping.

The investigations that can be carried out with state-of-the-art simulators such as \emph{Synthetic LISA} include:

\mylistitem{Performance characterization and architecture trade-off studies.}
Synthetic time series supplement analytical results in the allocation of subsystem noise budgets and in the determination of the final sensitivity for specific GW sources, providing a high-level analysis tool for system engineering, and helping the formulation of technical requirements from the desired LISA science goals.

An example was the recent study \cite{gair04} of detection prospects for the GW signals from compact stellar objects inspiraling into the supermassive black holes at the centers of galaxies, with the purpose of determining whether the LISA noise floor would need to be lowered to guarantee a minimum number of such detections. For this study, time series for $h_+$ and $h_\times$ were produced using the Glampedakis--Hughes--Kennefick quasiadiabatic orbit integrator \cite{ghk2002}, and then fed to \emph{Synthetic LISA}, which computed the corresponding time series of TDI observables; these were used to derive the expected S/Ns for the capture sources.

\mylistitem{Noise analysis and vetos.} Synthetic time series can be used to study real-LISA features of the instrumental noises, such as nonstationarity, noise increments due to faulty subsystems, or (perhaps most important) the level of cancellation of laser phase noise by TDI under different LISA geometries, armlength-measurement tolerances, and other TDI characteristics. The numerical experiments presented in this paper represent a first step in the numerical validation of TDI as implemented for LISA; more detailed studies will undoubtedly become necessary as additional details about the actual implementation of TDI become available.

A recent example was the use of \emph{Synthetic LISA} \cite{tdir} to validate a new approach to the determination of the LISA armlengths, whereby the noise power in the TDI observables is minimized as a function of the armlengths.

\mylistitem{Development of data-analysis algorithms.} The synthetic time series produced by this simulation have consistent signal structure and noise correlations across all the TDI combinations. Thus they can be used to test algorithms for use on the real LISA data, such as the separation of stochastic GW backgrounds from LISA instrumental noises \cite{TAE2000}, the matched-filtering detection of quasiperiodic signals \cite{ktv}, and so on. \emph{Synthetic LISA} provides a streamlined module to filter GWs through the LISA TDI response, allowing easy interfacing to existing GW data or GW-modeling applications. GW data analysts using \emph{Synthetic LISA} to generate simulated LISA data will also be able to exploit the library of GW signals being assembled at the \emph{Mock LISA data archive} \cite{mock}.
\lastlistitem

\noindent \textbf{Acknowledgements.}
The author wishes to thank John Armstrong, Frank Estabrook, and Massimo Tinto for teaching him about all things TDI; and also Tom Prince, Bonny Schumaker, Andrzej Kr\'olak, Jeff Edlund, Daniel Shaddock, Bob Spero, Brent Ware, and Neil Cornish for useful discussions and interactions.  John Armstrong developed precursor software to \emph{Synthetic LISA} and assisted the development and testing of the current version, while Jeff Edlund provided help with coding.  This research was supported by the LISA Mission Science Office at the Jet Propulsion Laboratory, California Institute of Technology, where it was performed under contract with the National Aeronautics and Space Administration.
\begin{figure}\begin{center}
\includegraphics[width=3.2in]{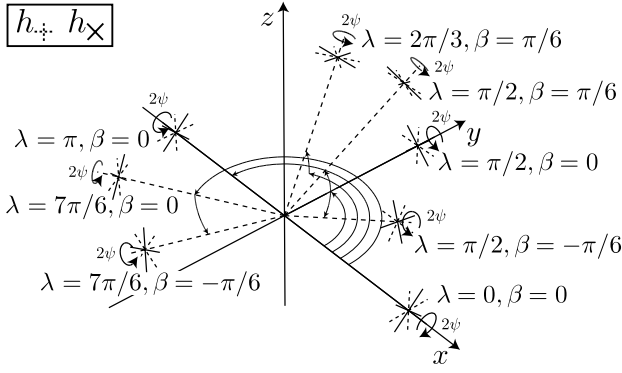}
\end{center}\caption{Conventional definition of the GW polarizations $+$ (dashed) and $\times$ (solid) for various ecliptic latitudes $\beta$ and longitudes $\lambda$. Figure excerpted from Ref.\ \cite{ktv}.\label{fig:polconv}}
\end{figure}

\appendix

\section{Geometric conventions in \emph{Synthetic LISA}}
\label{app:conventions}

In this Appendix we give a brief overview of the \emph{Synthetic LISA} conventions for the LISA orbits and for the geometry of GW sources, with the purpose of facilitating the comparison between the synthetic signals generated by \emph{Synthetic LISA}, and those created with other methods (and in particular with the \emph{LISA Simulator} \cite{cornishrubbo}). As discussed in Sec.\ \ref{sec:lisamodel}, the motion of the LISA array baricenter is approximately contained in the plane of the ecliptic; for this and other reasons \cite{TEA2004}, \emph{Synthetic LISA} employs a Solar-system--baricentric ecliptic coordinate system (SSB frame), setting the $x$ axis toward the vernal point.

The most accurate description of the LISA motion available in \emph{Synthetic LISA} (the \texttt{LISA} class \texttt{EccentricInclined}) models the eccentric orbits of the spacecraft up to second order in the eccentricity $e$. For these orbits, the dominant (and evolving) differences between the armlengths are caused by the flexing motion of the array \cite{flexy2003} due to orbital eccentricities. Following Ref.\ \cite{cornishrubbo}, the SSB coordinates $\vect{p}_i$ of spacecraft $i$ are given by
\begin{widetext}
\begin{equation}
\left[\begin{array}{c}
p_i^x \\ p_i^y \\ p_i^z
\end{array}\right] = (1\,\mathrm{AU})
\left[\begin{array}{c}
\cos \alpha + e [\sin \alpha \cos \alpha \sin \beta_i - (1 + \sin^2 \alpha) \cos \beta_i] + O(e^2) \\
\sin \alpha + e [\sin \alpha \cos \alpha \cos \beta_i - (1 + \cos^2 \alpha) \sin \beta_i] + O(e^2) \\
-\sqrt{3} e \cos(\alpha - \beta) + O(e^2)
\end{array}\right],
\label{eq:eccentricinclined}
\end{equation}
\begin{equation}
\alpha = \Omega t + \eta_0, \quad
\beta_i = \eta_0 + \xi_0 - \sigma_i,
\label{eq:eccentricinclinedb}
\end{equation}
\end{widetext}
where $\sigma_i = 3\pi/2 - 2(i-1)\pi/3$ and $e = 0.00964838$, yielding an effective $L \simeq 16.6782$ s. These spacecraft orbits are mapped to those used in the \emph{LISA Simulator} \cite{cornishrubbo} by setting $\eta_0 = \kappa$, $\xi_0 = 3 \pi / 2 - \kappa + \lambda$, where $\kappa$ and $\lambda$ are the parameters defined below Eqs.\ (56) and (57) of Ref.\ \cite{cornishrubbo}, and by choosing $\mathit{sw} < 0$ in the \texttt{EccentricInclined} constructor, which has the effect of exchanging spacecraft 2 and 3.

The armlengths experienced by light propagating along the arms can be found by solving Eq.\ \eqref{eq:lightprop}. For efficiency, \emph{Synthetic LISA} employs the lowest-order approximation
\begin{multline}
\label{eq:flexyarms}
L_\mathit{arm} = L + \frac{1}{32} (e L) \sin(3\Omega t - 3\xi_0)
+ \left[ (\mathrm{sgn}\,\mathit{arm}) (\Omega R L) \right. \\
- \left. \frac{15}{32} (e L) \right] \sin(\Omega t - \delta_{|\mathit{arm}|}),
\end{multline}
where $\delta_i \equiv \{\xi_0,\xi_0 + 4\pi/3,\xi_0 + 2\pi/3\}$. The amplitude of the flexing correction is about $7.5 \times 10^{-2}$ s, or 0.5\% of the nominal LISA armlength; the rate of change of the armlengths is about $1.5 \times 10^{-8}$ s/s, which requires second-generation TDI to yield sufficient cancellation of laser phase noise.

All the \emph{Synthetic LISA} GW source objects (\texttt{Wave}) share the same geometrical setup, which follows the conventions of Ref.\ \cite{ktv}. At the position $\vect{x}$ in the SSB frame, the spatial part of the transverse--traceless metric perturbation associated with a plane GW can be written as
\begin{equation}
\tens{h}(t) = 
h_+(t - \hat{k}\cdot\vect{x}) \, \tens{e}_+ +
h_\times(t - \hat{k}\cdot\vect{x}) \, \tens{e}_\times;
\label{eq:eqpol}
\end{equation}
here the functions $h_+(t)$ and $h_\times(t)$ express the two
polarization components of the wave at time $t$, measured at the origin of
the SSB frame. For a GW source at ecliptic latitude $\beta$ and ecliptic longitude $\lambda$, the unit propagation vector $\hat{k}$ is
\begin{equation}
\label{eq:kvec}
\hat{k} \equiv -(\cos \beta \cos \lambda, \cos \beta \sin \lambda,
\sin \beta).
\end{equation}
The two polarization tensors $\tens{e}_+$ and
$\tens{e}_\times$ that appear in Eq.\ \eqref{eq:eqpol} are defined without loss of generality as
\begin{equation}
\tens{e}_+ \equiv \tens{E} \cdot \left( \begin{array}{ccc} 1 & 0 & 0 \\ 0 & -1 & 0 \\ 0 & 0 &
0 \end{array} \right) \cdot \tens{E}^T, \quad
\tens{e}_\times \equiv \tens{E} \cdot \left( \begin{array}{ccc} 0 & 1 & 0 \\ 1 & 0 &
0 \\ 0 & 0 & 0 \end{array} \right) \cdot \tens{E}^T,
\label{eq:poltensors}
\end{equation}
where the orthogonal matrix $\tens{E}$,
\begin{widetext}
\begin{equation}
\tens{E} \equiv 
\left( \begin{array}{ccc}
 \sin \lambda \cos \psi  - \cos \lambda \sin \beta \sin \psi  &
- \sin \lambda \sin \psi -\cos \lambda \sin \beta \cos \psi  &
-\cos \lambda \cos \beta  \\
-\cos \lambda \cos \psi  - \sin \lambda \sin \beta \sin \psi  &
\cos \lambda \sin \psi - \sin \lambda \sin \beta \cos \psi  &
-\sin \lambda \cos \beta  \\
\cos \beta \sin \psi  &
\cos \beta \cos \psi  &
-\sin \beta 
\end{array} \right),
\end{equation}
\end{widetext}
expresses an Euler rotation sequence, whereby the $\beta$ and $\lambda$ terms can be understood as enforcing the transversality of the GW, while the polarization angle $\psi$ encodes a rotation around the direction of wave propagation, $-\hat{k}$, setting the convention used to define the two polarizations. The polarizations corresponding to $\psi = 0$ are shown in Fig.\ \ref{fig:polconv} for various source positions in the sky.
The positional parameters $\beta$, $\lambda$, and $\psi$ are mapped to the parameters $\theta$, $\phi$, and $\psi$ used in the \emph{LISA Simulator} \cite{cornishrubbo} by setting $\beta = \pi/2 - \theta$, $\lambda = \phi$, and $\psi = -\psi$.

\begin{figure}\begin{center}
\includegraphics[width=3.0in]{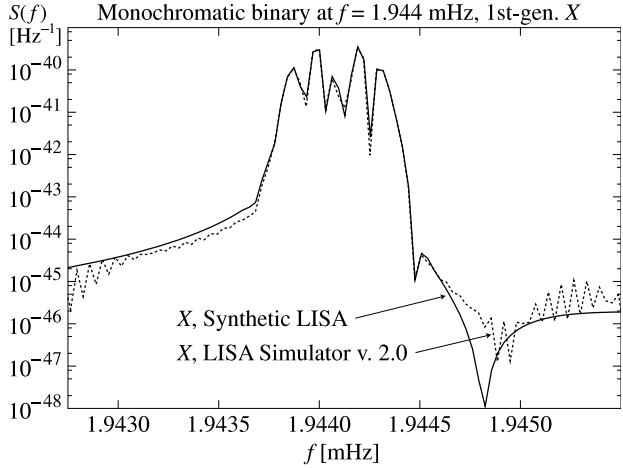}
\end{center}\caption{
Comparison between the outputs of the \emph{LISA Simulator}, v.\ 2.0 \cite{cornishrubbo} using \cpp{Newtonian.c} and of \emph{Synthetic LISA} for the response of first-generation TDI $X$ to a monochromatic binary with $f = 1.944$ mHz, $\beta = \lambda = \psi = 0$, $\iota \simeq \pi/2$, and amplitude appropriate for a $(0.5 + 0.033)M_\odot$ binary at 0.1 kpc.
We show the nonaveraged spectra of the unwindowed signals in a neighborhood of the central frequency. To compare the fractional-frequency-fluctuation output of \emph{Synthetic LISA} to the nominal-strain output of the \emph{LISA Simulator}, we have multiplied the strain spectrum by the square of the nominal armlength ($10^{10}$ m) to get a displacement spectrum; converted to a velocity spectrum using the derivative theorem for Fourier transforms; and converted to a fractional-frequency-fluctuation spectrum by dividing by $c^2$.\label{fig:test-binary}}
\end{figure}

The standard monochromatic-binary \texttt{Wave} object, \cpp{SimpleBinary} implements the GW signal
\begin{equation}
\left[\begin{array}{c}
h_+(t) \\
h_\times(t)
\end{array}\right] =
A
\left[\begin{array}{r@{\,}l}
(1 + \cos^2 \iota) & \times \cos(2 \pi f t + \phi_0) \\
(2 \cos \iota) & \times \sin(2 \pi f t + \phi_0)
\end{array}\right],
\end{equation}
where $A$ is the common amplitude, $\iota$ is the inclination angle, $f$ is the GW frequency observed in the SSB frame, and $\phi_0$ is the phase at $t=0$. The standard value of $A$ is $(2 m_1 m_2 / d \, R)$ with $m_1$, $m_2$ the two masses, $d$ the luminosity distance, and $R$ the orbital separation (the common amplitude $h_0$ used in Ref.\ \cite{ktv} differs by a factor of two, $h_0 = 2 A$, absorbed in $h_0^\times$ and $h_0^+$). We have found excellent agreement (see, e.g., Fig.\ \ref{fig:test-binary}) between the time series of TDI observables derived from \cpp{SimpleBinary} and the output of the \emph{LISA Simulator}, v.\ 2.0 \cite{cornishrubbo} using \cpp{Newtonian.c}.

\section{Synthetic Noise in \emph{Synthetic LISA}}
\label{app:synthnoise}

In \textit{Synthetic LISA}, pseudorandom white noise is created by generating a sequence of uncorrelated Gaussian deviates,\footnote{Independent, uniformly distributed deviates are obtained from Luescher's lagged Fibonacci generator \cite{luescher}, as implemented in the \emph{GNU Scientific Library} \cite{gsl}; the Box--Muller transform \cite{boxmuller} is then used to convert the uniform deviates to Gaussian deviates.} which are then interpreted as the sampled values at times $t_n = n \Delta t$ (for $n = 0, 1, \ldots$) of a continuous random process. The process is assumed to be bandlimited below $f_b = 1/(2 \Delta t)$: by the sampling theorem (see, e.g., Ref.\ \cite{nrc}), the value of the noise can then be reconstructed exactly at any intervening time $t$ by convolving the sampled sequence with the interpolating kernel
\begin{equation}
\mathrm{sinc}[\pi (t - t_n) / \Delta t] =
\frac{\sin[\pi (t - t_n) / \Delta t]}{[\pi (t - t_n) / \Delta t]}. 
\end{equation}
Since the sinc kernel has infinite time extent, it must be replaced in practice by an approximated interpolation scheme that involves a finite number of samples. A vast class of such schemes, including the linear and polynomial interpolators implemented in \emph{Synthetic LISA}, can be formulated as the convolution of the sampled sequence with an interpolating kernel that is (in some sense) an approximation to the sinc.
\begin{figure}\begin{center}
\includegraphics[width=3.3in]{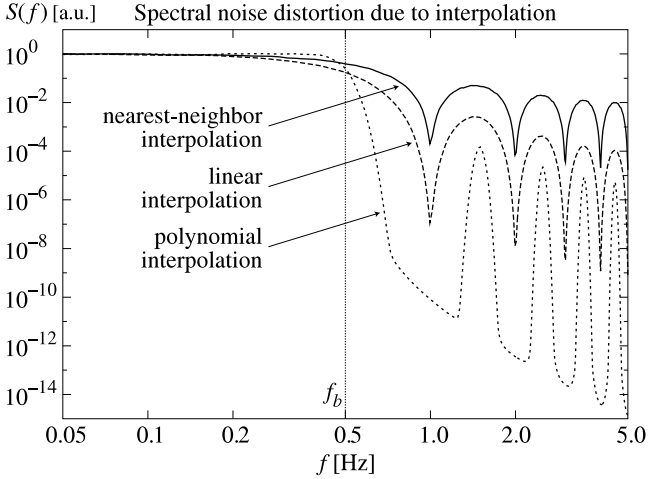}
\end{center}\caption{Noise distortion for 0.5 Hz-bandlimited white noise, with different interpolation schemes. Higher-order interpolation corresponds to a sharper transition at the bandlimit frequency and to lower ripples and deeper valleys between them (in this graph, the valleys are cut off by spectral leakage from the main platform at power $= 1$).\label{fig:interpolation}}
\end{figure}

The tradeoff in the approximation is between the number of samples used to interpolate and the sharpness of the spectral response. The correct sampling of a bandlimited process preserves all the spectral information below the Nyquist frequency, but it populates Fourier space with infinitely many replicas of the original spectrum, centered at frequencies $k / \Delta t$ (for $k = \pm 1, \pm 2, \ldots$). The effect of sinc interpolation is to multiply this composite spectrum by the Fourier transform of the sinc, which is a perfect square window of height 1 and width $1 / \Delta t$, centered at $f = 0$. Thus, sinc interpolation achieves perfect signal reconstruction by selecting only the original spectrum and deleting all unwanted replicas. Practical schemes with kernels of finite extent cannot have such a sharp frequency response, so they distort (i.e., amplify or suppress, depending on frequency) the original spectral content in the passband below $f_b$, and they allow some of the power of the unwanted spectral replicas to creep back into the interpolated process (either directly, if the process is sampled with a sufficiently high Nyquist frequency, or by aliasing to frequencies in the passband).

These effects can be observed in Fig.\ \ref{fig:interpolation}, which shows a spectrum of pseudorandom white noise, generated with a timestep of 1 s, and resampled to a timestep of 0.1 s, using no interpolation (i.e., defaulting to the nearest 1-s sample), using linear interpolation, and using Lagrange-polynomial interpolation of order 4, 8, and 32. In all cases, power begins to drop before the nominal bandlimit frequency of 0.5 Hz, but the drop is sharper and closer to 0.5 Hz for higher-order interpolation methods. Spurious power above the bandlimit frequency appears as \emph{ripples} between the $f_b$ harmonics: the height of the ripples decreases with the interpolation order, while the valleys among the ripples become wider. In Fig.\ \ref{fig:interpolation}, the valleys appear to be cut off by a common downgrading envelope; this is an artifact of spectral estimation, due to the residual leakage from the platform below the passband; spectral leakage also smears out to a finite height the nulls at the $f_b$ harmonics.

In \emph{Synthetic LISA}, interpolated pseudorandom white noise is used to stand in for the standard laser phase noise of Sec.\ \ref{sec:noiseresponse}. The standard proof-mass and optical-path noises, which have colored spectra, are approximated by applying simple digital time-domain filters to the uncorrelated deviates, before interpolation. Namely, the finite-difference filter $y[n \Delta t] = x[n \Delta t] - x[(n-1) \Delta t]$ (with $x$ the original noise sequence) has power transfer function $|1 - \exp(2 \pi i f \Delta t)|^2 = 4 \sin^2 (\pi f \Delta t)$, and is used to approximate the standard $f^{-2}$ proof-mass noise. The damped-integrator filter $y[n \Delta t] = \alpha y[(n-1) \Delta t] + x[n \Delta t]$ (with $\alpha = 0.9999$, to control the DC component of $y$) has power transfer function $\simeq (1/4) \sin^{-2} (\pi f \Delta t)$, and is used to approximate the standard $f^2$ optical-path noise.

The resulting pseudorandom noises have power spectra that adhere very faithfully to the nominal curves, except at frequencies comparable to $f_b$, where the effect of interpolation is that noise power is not cut off sharply, but rather drops off smoothly (if rapidly), with nulls at the $f_b$ harmonics. For the optical-path and proof-mass noises, the effect of interpolation is compounded by the effect of the finite-difference and finite-integration time-domain filters, whose transfer function near $f_b$ is proportional to $\sin^{\pm 2} [\pi f / (2 f_b)]$ rather than $f^{\pm 2}$. We conclude that the pseudorandom noises can be accurate representations of the standard LISA noises of Sec.\ \ref{sec:noiseresponse}, and therefore can be used to study the noise response of the TDI observables, as long as we take into account the effects of interpolation and filtering at frequencies comparable to $f_b$. Because TDI is essentially a linear operation, the results at lower frequencies will not be affected. Using linear interpolation (the \emph{Synthetic LISA} default), it is probably safe to draw conclusions from the TDI results at frequencies $\lesssim f_b/5$; using higher-order interpolation, it becomes possible to push inferences to higher frequencies.

This discussion of filtering and interpolation applies also to noise objects provided by the user as sampled time series, as long as the sampled noise can be considered bandlimited below its nominal Nyquist frequency. See Ref.\ \cite{lisainterpolation} for a related discussion of the use of interpolation in reconstructing the TDI observables on Earth from the $y_{slr}$ and $z_{slr}$ data, sampled onboard at a limited rate that can be transmitted affordably to Earth.

\end{document}